\documentclass[authorversion,nonacm]{acmart}
\AtBeginDocument{%
  \providecommand\BibTeX{{%
    \normalfont B\kern-0.5em{\scshape i\kern-0.25em b}\kern-0.8em\TeX}}}

\setcopyright{none}
\copyrightyear{2024}
\acmYear{2024}
\acmDOI{XXXXXXX.XXXXXXX}

\usepackage{float}
\usepackage{xspace}
\usepackage{graphicx}
\usepackage{tabularx}
\usepackage{svg}
\usepackage{wrapfig}
\usepackage{subcaption}

\newcommand{\sayit}[1]{``\textit{#1}''}
\newcommand{\say}[1]{``#1''}

\newboolean{showcomments} 
\setboolean{showcomments}{false}
\ifthenelse{\boolean{showcomments}}
{\newcommand{\nb}[2]{
		\fbox{\bfseries\sffamily\scriptsize#1}
		{\sf\small$\blacktriangleright$\textit{\textcolor{blue}{#2}}$\blacktriangleleft$}
	}
	
}
{\newcommand{\nb}[2]{}}

\newcommand\ignore[1]{}

\newcommand\chenglong[1]{{}}
\newcommand\jeevana[1]{{}}

\newcommand{\tool}{\textsc{DynaVis}\xspace}
\newcommand{\vgl}{Vega-Lite\xspace}
\newcommand{\control}{$\mathbf{UI_{NL}}$\xspace}
\newcommand{\experiment}{$\mathbf{UI_{DW}}$\xspace}

\begin{document}

\title{\tool: Dynamically Synthesized UI Widgets for Visualization Editing}

\author{Priyan Vaithilingam}
\email{pvaithilingam@g.harvard.edu}
\affiliation{%
  \institution{Harvard University}
  \city{Boston}
  \country{USA}
}

\author{Elena L. Glassman}
\email{glassman@seas.harvard.edu}
\affiliation{%
  \institution{Harvard University}
  \city{Boston}
  \country{USA}}

\author{Jeevana Priya Inala}
\email{jinala@microsoft.com}
\affiliation{%
  \institution{Microsoft}
  \city{Redmond}
  \country{USA}
}

\author{Chenglong Wang}
\email{chenwang@microsoft.com}
\affiliation{%
 \institution{Microsoft}
 \city{Redmond}
 \country{USA}}

\renewcommand{\shortauthors}{Vaithilingam, et al.}

\begin{abstract}
Users often rely on GUIs to edit and interact with visualizations --- a daunting task due to the large space of editing options. As a result, users are either overwhelmed by a complex UI or constrained by a custom UI with a tailored, fixed subset of options with limited editing flexibility.  Natural Language Interfaces (NLIs) are emerging as a feasible alternative for users to specify edits. However, NLIs forgo the advantages of traditional GUI: the ability to explore and repeat edits and see instant visual feedback.

We introduce \tool, which blends natural language and dynamically synthesized UI widgets. As the user describes an editing task in natural language, \tool performs the edit and synthesizes a persistent widget that the user can interact with to make further modifications. Study participants (n=$24$) preferred \tool over the NLI-only interface citing ease of further edits and editing confidence due to immediate visual feedback.
\end{abstract}

\begin{teaserfigure}
  \centering
  \includegraphics[width=\linewidth]{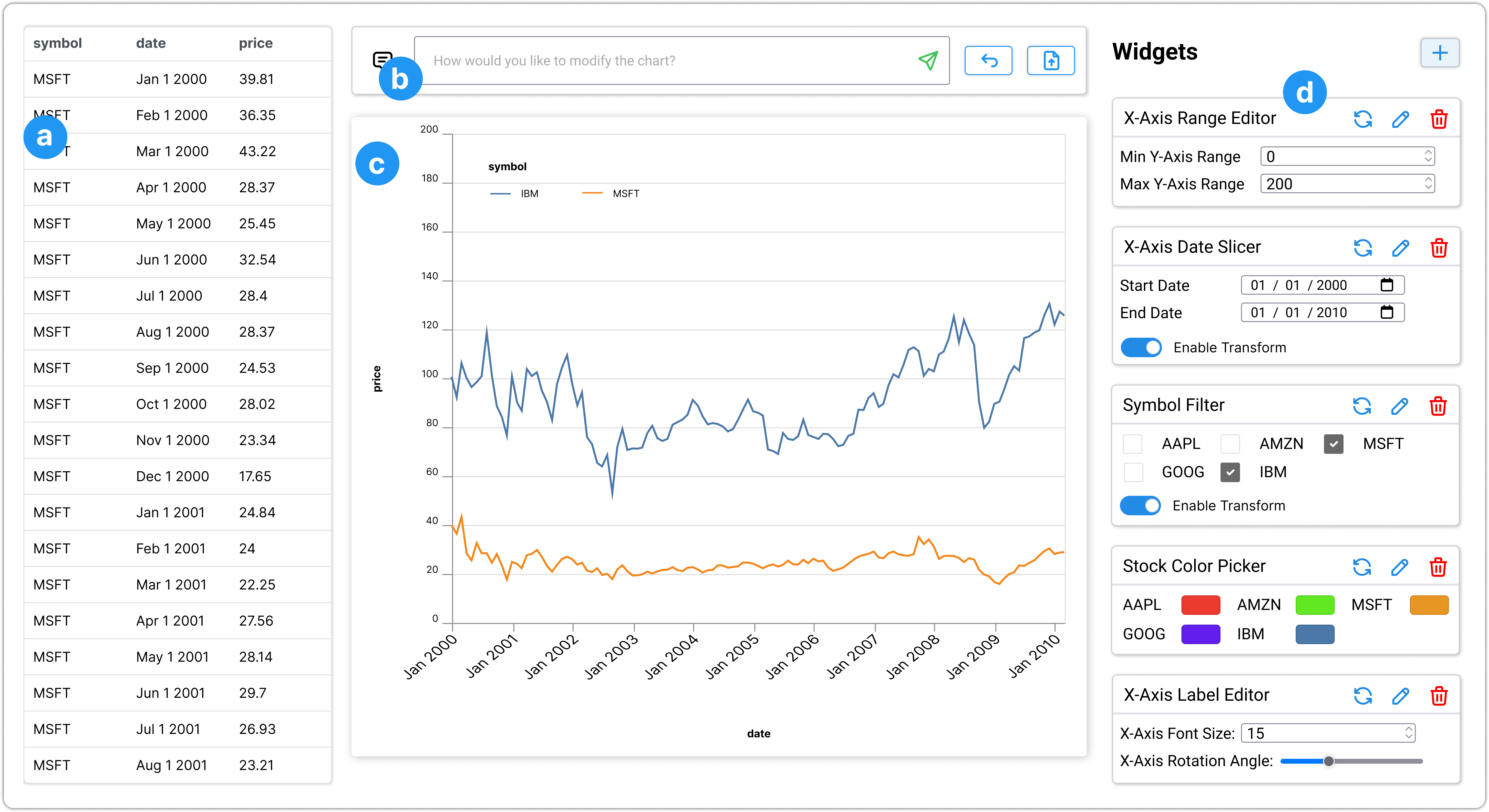}
  \caption{Screenshot of \tool tool. (a) Imported data is shown on the left as a table. (b) Users can provide natural language command to edit the chart using the command bar. (c) The visualization is displayed on the center. (d) The widgets panel shows the automatically synthesized dynamic widgets based on user' natural language commands, in reverse chronological order (recently added widgets at the top).}
  \Description{\tool in action}
  \label{fig:ui}
\end{teaserfigure}

\maketitle

\section{Introduction}

Modern interactive visualization authoring tools (e.g., Tableau~\cite{tableau}, PowerBI~\cite{powerbi-home}, Lyra~\cite{satyanarayan2014lyra}, Charticulator~\cite{ren2018charticulator}) 
have greatly reduced the effort to create initial visualizations from data. With these tools, authors only need to specify high-level mappings from data fields to visual properties, and behind the scenes, these tools automatically provide ``smart defaults''~\cite{wilkinson2012grammar, satyanarayan2016vega} to fill in hundreds of chart parameters---hiding low-level details. 

\begin{figure}
    \centering
    \includegraphics[width=\linewidth]{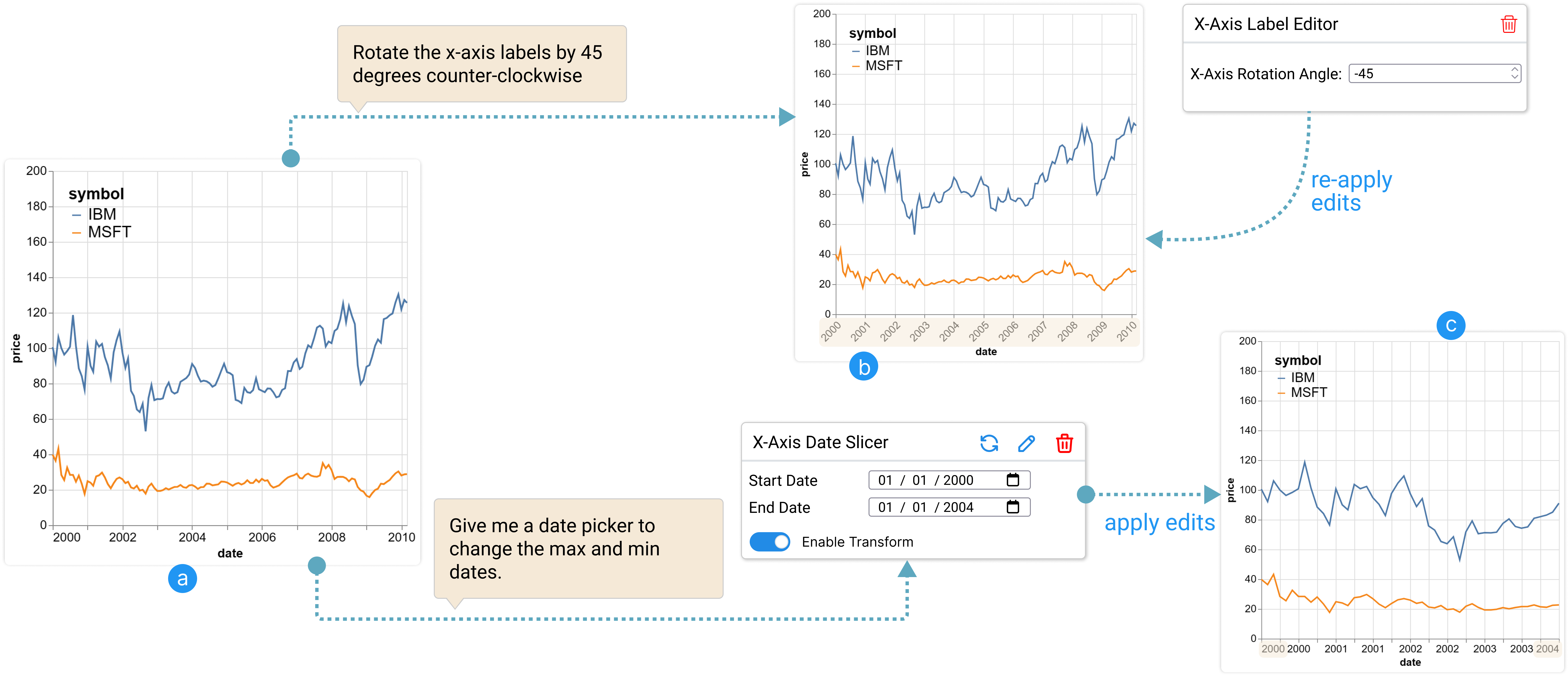}
    \caption{\tool dynamically synthesizes widgets based on natural language commands for visualization editing. The user can describe an edit to the visualization, and \tool modifies the visualization and synthesizes a dynamic widget which the user can use for further edits (shown as $a \rightarrow b$). Alternatively, the user can directly ask for a dynamic widget to perform edits (shown as $a \rightarrow c$).}
    \label{fig:intro}
\end{figure}

While these smart defaults are often sufficient for exploratory analysis, authors who want to refine the visualization to better communicate their insights and readers who want to customize the visualization to answer their analysis objectives often find themselves in need of \emph{editing} these default visualizations. For example, to prevent longer labels from overlapping in a line chart, the user has to rotate the labels in the $x$-axis. Or, the user will have to add a filter to only include data within a given date range (see Figure~\ref{fig:intro}).

These edits are often considered \say{small tweaks} of the visualization, but these long-tailed edits can be very challenging. First, the user needs to distinguish which options will lead to the desired editing effect (e.g., understand that they need the \say{tick} option as opposed to \say{scale} or \say{legend} to edit label angle), which requires expertise on low-level visualization grammar. Then, the user needs to discover the edit option in the tool which may be buried in tiers of menus and panels among all others in a tool GUI (e.g., the user needs to right-click the $x$-axis to open its property editor, locate the sub-panel on ticks, find the rotation option to change the label angle), which can be challenging to achieve without decent tool expertise. As a result, users are either presented with a complex UI where they are swamped with options, or a tailored interface designed to simplify navigation where they often find themselves too restricted to perform the desired customization.

An emerging approach to address this visualization editing and refinement challenge is to design \emph{natural language interfaces} (NLIs) that allow users to describe editing effects in natural language. Then, based on the user's instructions, the tools automatically infer necessary options and corresponding values to apply the edits. For example, the user can give the natural language command to \say{move the $x$-axis title to the left side of the axis}, which will translate to changing the \say{titleAnchor} property of the \say{$x$} encoding to the value \say{start}. However, while NLIs address the discovery and navigation challenges, they forgo the benefits of GUI, especially the abilities to {perform fine-grained edits}, {obtain immediate visual feedback from editing results}, and {quickly undo and reapply edits}. For example, if the user wants to make the width of the strokes in the line chart thicker, they do not always have the exact size in mind, and would often try out different sizes before choosing one. Or, if the user wants to change the colors of the bar chart, they may not know the exact hex (or RGB) value to provide. 
Such limitations restrict NLIs' applications in visualization editing.

To address the visualization editing and refinement challenge, we design a new interaction approach, \emph{interaction via dynamically synthesized GUI widgets}, and develop a tool named \tool for visualization editing. Our key design insight is to blend natural language interfaces with interactive GUI widgets so that users can benefit from both NLIs' reduction in the gulf of execution~\cite{nn-gulf} and GUI's interactivity. To perform a visualization editing task, the user starts by either describing what edits they want to perform (e.g., ``rotate $x$-axis label 45 degrees'') or directly asking for a GUI widget that they envision to perform the edits with (e.g., ``give me a slider to control $x$-axis label angles''); either way, \tool synthesizes a GUI widget (along with a preset value from user's specification in the former case) using a Large Language Model (LLM) for the user to explore and perform subsequent edits. 

Besides the immediate benefits of reduced navigation overhead and interactivity for exploring edit effects, users can also easily compose and coordinate multiple edits using dynamic widgets as they persist after synthesis for quick editing access. Behind the scenes, we designed a widget synthesis engine powered by a large language model that translates user inputs into an HTML implementation of the widget and a call-back function that connects the widget inputs to visualization properties. \tool is highly expressive and supports both chart design edits (e.g., adjusting tick spacing, legend position, color scheme, label and title font properties) for authors to refine visualizations, and data-related edits (e.g., generating filters, zooming controllers, and sort) for readers to interactively explore visualizations without pre-built interactive widgets. Our study with 24 participants shows that participants prefer to use \tool over NLI-only interfaces due to the ease of repeating edits, and increased confidence when editing using a GUI due to immediate visual feedback.

Our contributions are as follows:
\begin{itemize}
    \item A new interactive approach for visualization editing, dynamic widgets, that combines NLI with GUI widgets to reduce the gulf of execution and enhance interactivity.
    \item A widget synthesis engine that leverages large language models to translate natural language inputs into widgets and control functions.
    \item A user study to evaluate how users use \tool to solve visualization editing tasks.
\end{itemize}

\section{Usage Scenario}~\label{sec:scenario}

\begin{figure}[h]
    \centering
    \includegraphics[width=\linewidth]{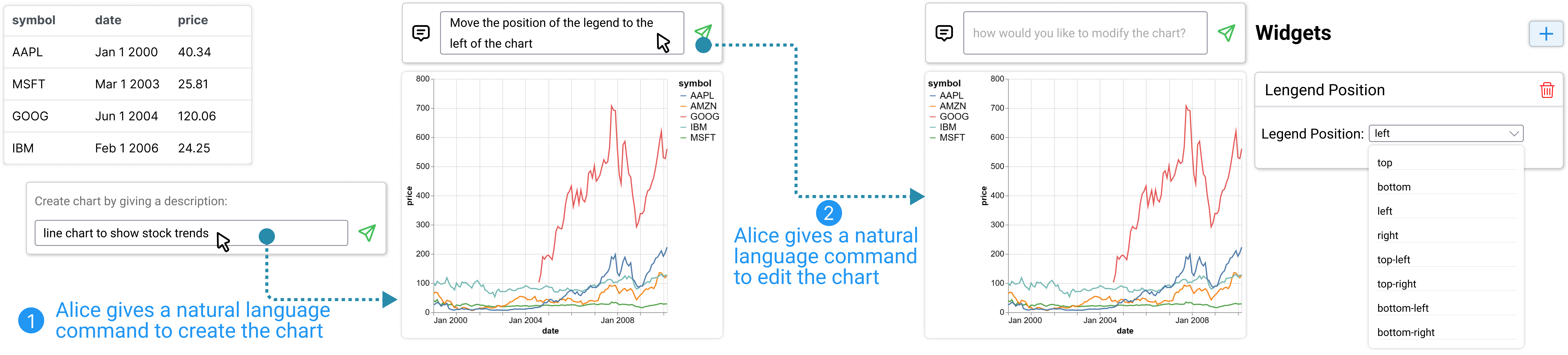}
    \caption{(1) Alice asks \tool to create a line chart to show the stock trend by providing a natural language prompt. (2) She then asks \tool to edit the legend position by giving a natural language command}
    \label{fig:process1}
\end{figure}

Alice is a consultant analyzing stock trends of technology companies using spreadsheets, and she needs to create visualizations to present her analysis results to her collaborator. Below, we describe Alice's experience of using \tool to edit and enhance her charts. 
\autoref{fig:ui} shows the UI of \tool, which contains four main components: (a) the data panel, (b) the command bar for specifying visualization and editing commands, (c) the visualization panel that shows the current working chart, and (d) the panel of synthesized dynamic widgets that users can use to manage widgets and edit the working chart. 

\paragraph{\bf{Initial chart.}}
Alice starts by importing the data \say{stocks.csv} into the tool, and the data shows up in the data panel (\autoref{fig:ui}-a). To create a chart, she provides a natural language description of the chart \sayit{create a line chart showing the stock trends} in \autoref{fig:ui}-b. Upon submission, \tool invokes an LLM to generate a line chart based on information from the dataset and the NL description (\autoref{fig:process1}). Besides creating the chart using natural language, Alice can also import the Vega-Lite visualization spec she created from other tools. 

Alice is not quite satisfied with the initial visualization because (1) the legend takes too much space on the right, (2) $x$-axis labels are too small to read, and (3) the color scheme is not ideal, Alice decides to use \tool to refine the chart. With \tool, Alice has two options to edit charts: (1) provide a natural language instruction in the command bar (\autoref{fig:ui}-b) to describe the edit she wants to achieve, and (2) explicitly add a widget by clicking the \say{\textsf{+}} button at the top of the widgets panel (\autoref{fig:ui}-d) and providing a natural language description of the desired widget. Either way, \tool dynamically generates widgets for Alice to perform edits.

\paragraph{\bf{Adjusting legend position via chart editing commands.}}
Alice first wants to adjust the legend position to keep the legend contained within the main chart canvas. She decides to use \emph{chart editing commands} to describe changes she wants to apply. For this, she provides the instruction \sayit{move the legend to the left of the chart} through the natural language command bar. Based on the instruction, \tool updates the visualization spec to re-position the legend. Additionally, \tool also automatically generates a widget with a drop-down menu for various legend positions pre-populated. As shown in \autoref{fig:process1}-(2), with this widget, Alice experiments with multiple legend positions before finalizing her final choice of \say{top-left corner}, which is, in fact, a better option than \say{left} that Alice didn't expect in the beginning. 

\begin{figure}[h]
    \centering
    \includegraphics[width=\linewidth]{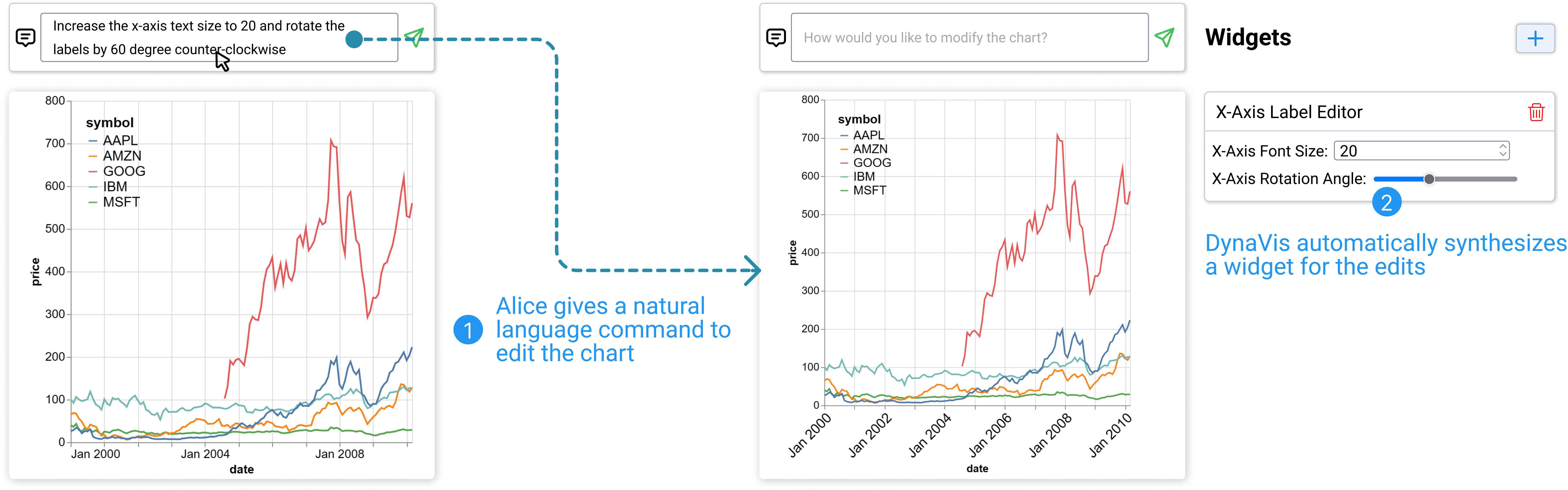}
    \caption{(1) Alice asks \tool to increase the $x$-axis label text size to 20 and rotate it by 45 degrees counter-clockwise. (2) \tool edits the chart and also adds a dynamically synthesized widget for Alice to make further changes in the future.}
    \label{fig:process2}
\end{figure}

\paragraph{\bf{Coordinated editing of text size and rotation angle of $x$-axis labels.}}
Next, Alice wants to resolve issues with $x$-axis labels which are currently too small to read. However, this can be a quite challenging edit: increasing the font size would most likely make labels overlap with each other; and while overlaps can be resolved by rotating label angles, too much rotation would make them less readable. Thus, Alice needs to coordinate the edits to label font size and rotation angle to find the right balance. Alice doesn't have a clear idea on what's the optimal combination yet, thus she starts with an exploratory command \sayit{increase the $x$-axis text size to $20$ and rotate the labels by $60$ degrees}. \tool updates the chart based on the command and presents her with a \textsf{\say{$x$-axis Label Editor}} widget that lets her edit the text size and angles of the $x$-axis labels simultaneously (see Figure~\ref{fig:process2}). With this, Alice can try out many combinations with ease without having to re-issue the editing instructions. After some trial-and-error, she settles on a font size of $15$ and rotation angle of $-45$ that suits her needs.

\begin{figure}[h]
    \centering
    \includegraphics[width=\linewidth]{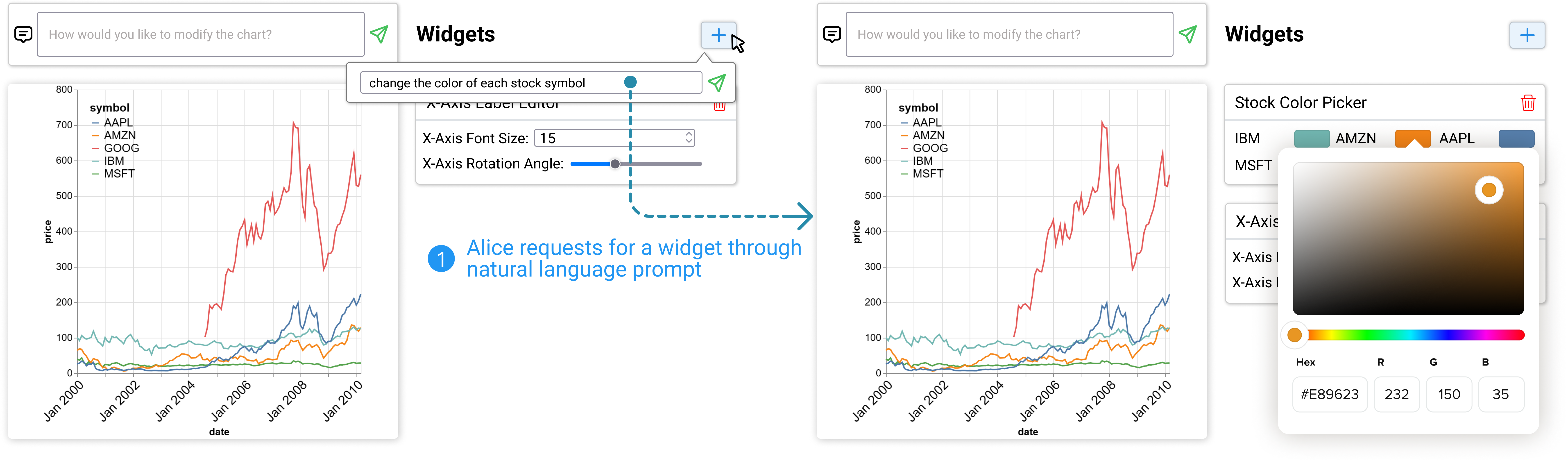}
    \caption{(1) This time Alice explicitly requests for a widget to change the color of each stock symbol by clicking the \say{\textsf{+}} button and giving the natural language prompt. (2) \tool synthesizes a widget, without making any edits to the chart.}
    \label{fig:process3}
\end{figure}

\paragraph{\bf{Choosing colors via widget creation commands.}}
Now, Alice wants to modify the default color palette of the chart. Since Alice does not yet have concrete colors in mind, she decides to use the widget creation feature of \tool to ask for a widget to explore the color options. To do so, she clicks the \say{\textsf{+}} button at the top of the widget panel (\autoref{fig:process3}) to open the widget creation command box, she then types in the prompt \sayit{change the color of each stock symbol}. \tool, which understands the current context of the chart and the current dataset, generates a tailored \textsf{\say{Stock color picker}} widget that allows her to pick colors for each stock symbol in the chart (see Figure~\ref{fig:process3}). \tool even knows all the symbols that are part of her current dataset, and can generate a tailored widget for her. Alice can use this widget to try different colors, get instant visual feedback, and choose the desired colors for her chart.

\paragraph{\bf{Data exploration with data filter widgets.}}
After Alice finishes up the customization, she emails the visualization to her collaborator Alex who plans to include the visualization in a news article he is writing. After some analysis, Alex wants to compare the stock trends for just \textsf{MSFT} and \textsf{IBM} to get deeper insights. Instead of going back and asking Alice to do that, Alex imports the visualization spec in \tool to perform the edits (a JSON \vgl spec that contains data as elaborated more in \autoref{sec:system}). After importing, Alex provides the command \sayit{compare only MSFT and IBM}, and \tool quickly updates the chart to run a filter transformation to show only the data for \textsf{MSFT} and \textsf{IBM}. \tool also provides a \textsc{\say{Symbol Filter}} widget with a checkbox for each stock symbol, using which Alex can continue to compare different stock trends choosing one or more symbols to compare (\autoref{fig:ui}, third widget). \tool also intelligently identifies that this widget has a data-filtering transform and provides a switcher for users to enable or disable the transformation.

Next, Alex asks for a \textsf{Date slicer} widget to zoom the visualization by using a smaller time window (\autoref{fig:ui}, second widget). Since the stock prices of the two companies that are compared are much lower than other companies, Alex also requests a widget to slice the $y$-axis range. Now using the generated \textsf{$y$-axis range slicer}, Alex can zoom in and out of the range window to see minute price changes within the time window of his choosing ( \autoref{fig:ui}, first widget). After finding the desired visualization, Alex includes the final visualization in his article for publication.

\paragraph{Remark:} Alice and Alex can complete the visualization refinement and exploration tasks with ease thanks to the following benefits of \tool.
\begin{itemize}
\item In conventional GUI, Alice needs to navigate menus and panels to locate widgets to perform edits, which requires her proficiency in both GUI and visualization terminologies. \tool lowers this barrier by allowing Alice to obtain desired widgets via natural language descriptions.
\item Synthesized widgets allow Alice to perform fine-grained edits and obtain immediate visual feedback from editing results, which enables her to explore and coordinate editing options. This allows her to find optimal edits through trial and error for edits she previously didn't know precisely (e.g., color, rotation angle, and text). Alice won't be able to explore edit options easily with an NLI as it requires her to describe concrete parameter values and has a delayed specification-feedback cycle.
\item \tool is highly expressive and supports both chart refinement edits for the author Alice and data manipulation edits for the reader Alex. Without \tool, Alex would either have to interact with Alice every time he wants an updated version of the chart or ask Alice to create an interactive visualization which requires additional efforts to tailor options.
\item As dynamic widgets persist after creation and are fully compositional, Alice and Alex can go back and repeat edits (e.g., update $x$-axis range as analysis objective changes) or revert certain edits (e.g., undo data filtering). This can be challenging with an NLI as changes are non-compositional, and every update requires a new editing command. Or, with a conventional GUI, they would need to keep track of all edits they have done in order to repeat or redo edits.
\end{itemize}

\section{\tool System Design and Implementation}
\label{sec:system}
In this section, we first describe the design principles behind our core concept of widgets. We then describe our synthesis framework for dynamically generating these widgets. \tool is a cross-platform web application that is implemented with React and Typescript for the front-end user interface and Python for the back-end server.

\subsection{Dynamic Widgets}
\paragraph{\bf{Widgets as modular sub-components.}}
\begin{figure}[!bht]
    \centering
        \includegraphics[width=0.9\textwidth]{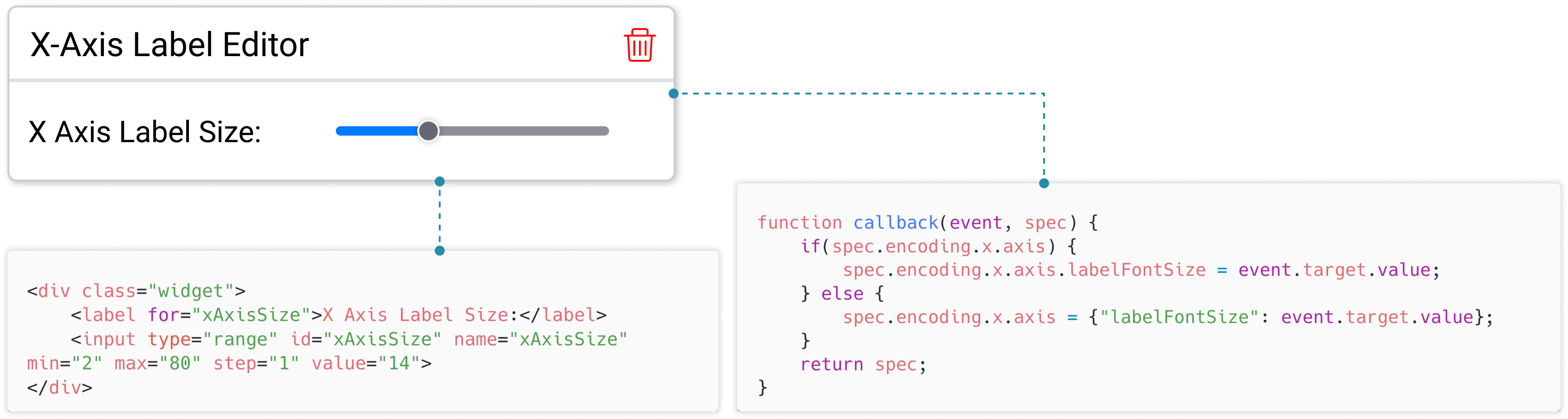}
    \caption{A dynamic widget is comprised of two components --- (1) The HTML script defining the UI (2) The JS callback function that listens for the changes in the UI to edit the visualization.}
    \label{fig:wid-1}
\end{figure}
With \tool, we introduce \textit{dynamic widgets}, which are small modular UI components that focus on a particular edit or interaction task at hand. The edit can involve simple changes to one or multiple chart properties, such as changing the position of the legend or slicing, or can involve a data transformation operation executed on the data before chart rendering, such as filtering specific ranges on the axes. At a high level, a widget has two components: 
\begin{itemize}
    \item An \emph{HTML script} that describes the UI elements of the widget which the user will interact with to manipulate the visualization.  e.g., An HTML script containing an \textsf{<input>} element of type slider to change the x-axis label size of the chart (see Figure~\ref{fig:wid-1}).
    \item A \emph{JavaScript callback function} that contains the code that will be executed to manipulate the visualization and/or data whenever the user interacts with the widget. This callback function is of the form \textsf{callback(event, chart) => (transforms, chart)}, which accepts both the HTML event object and the current chart as inputs and generates an optional list of transforms and the updated chart as outputs. This callback function is attached to the \textsf{onChange} event handlers of all the input elements in the HTML script.
\end{itemize}
An example of the dynamic widget is shown in Figure~\ref{fig:wid-1} along with its HTML script and Javascript callback function.

One of the main design choices of our system is to ensure the modularity of the widgets so that edits from two different widgets do not conflict or manipulate the chart in unexpected ways. This requirement has implications for how we handle charts and transforms: 

\paragraph{\bf{Handling charts.}} To ensure that each widget only changes a small component of the chart, we recognized that using a declarative chart representation is preferable to an imperative one. One such declarative representation is to use a JSON object (called a specification) to encode the properties of the chart. For example, with JSON representation, to change the x-axis title, one can just edit the chart specification as \textsf{spec.encoding.x.title = "axis title"} without having to change anything else regarding the chart. 

In this paper, we use 
\vgl specifications~\cite{satyanarayan2016vega} for representing charts since it provides us with a concise and declarative representation of visualization while also maintaining expressiveness. \vgl also provides a well-supported rendering engine that is compatible with all the major web frameworks.

\paragraph{\bf{Handling data transformations.}} Data transformations are an important part of visualization editing. \tool handles all the \vgl data transformations, e.g., filter, fold, flatten, etc.
In \tool, transforms are represented as a list of objects similar to \vgl. Each widget's callback outputs a list of transforms. The transforms are performed in the order in which they are specified in this list. If there are multiple widgets with transforms, we execute transforms in the chronological order of the widgets. To enable this, we keep a mapping of the widgets to their most recent callback's output transforms, so that we can execute all widget's latest transforms in the above order before rendering the chart for every edit. For every widget that adds transformations to the chart, \tool adds a switch so that users can enable or disable the transform. 
e.g., Let's take the example of Alex from Section~\ref{sec:scenario}. When they request a widget to slice the date range, the synthesized widget adds a \textit{filter transform} on the data. \tool identifies this as a special \textit{Transform Widget}, and allows Alex to dynamically enable or disable the transform to see the original chart and the filtered chart. 

\begin{figure}[t]
    \centering
    \includegraphics[width=\linewidth]{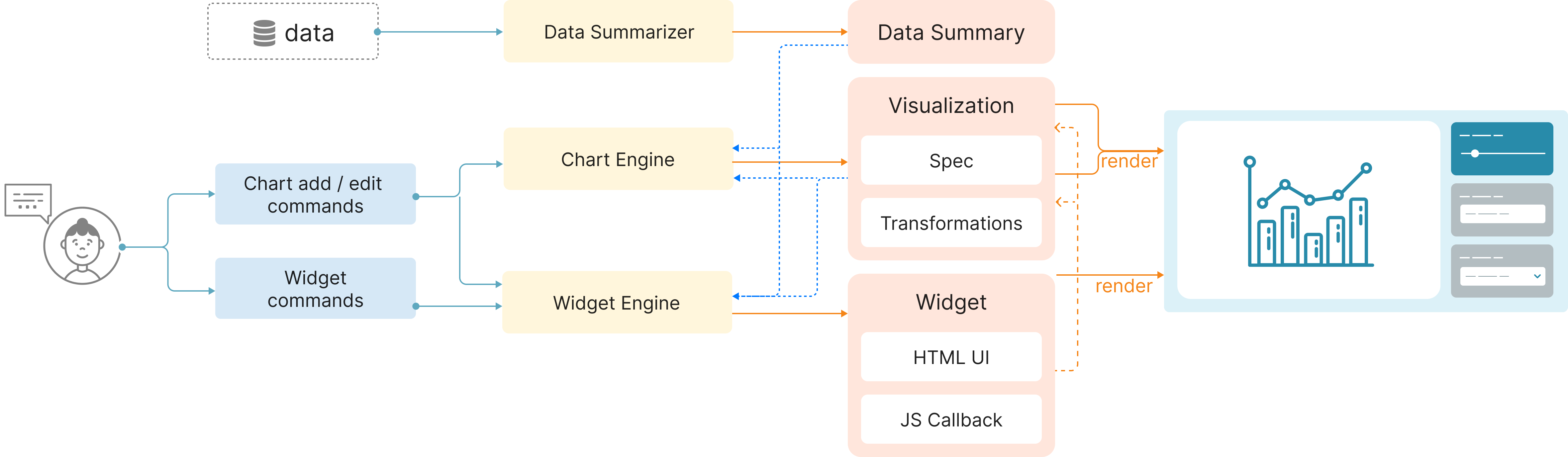}
    \caption{\tool system architecture. The Data Summarizer generates a summary from input data to assist visualization and widget synthesis. The Chart Engine applies changes to visualizations based on chart editing instructions, and the Widget Engine is responsible for synthesizing dynamic widgets from both types of user commands.}
    \label{fig:system}
\end{figure}
\subsection{Synthesis Framework}
Below we describe our synthesis framework by splitting it into three stages: pre-processing, LLM-based synthesis, and post-processing. 
Our framework comprises of 3 main modules: a Data Summarizer, a Chart Engine, and a Widget Engine as shown in  Figure~\ref{fig:system}. 

\begin{figure}[t]
\centering
\begin{minipage}[b][][b]{0.4\linewidth}
\includegraphics[width=\textwidth]{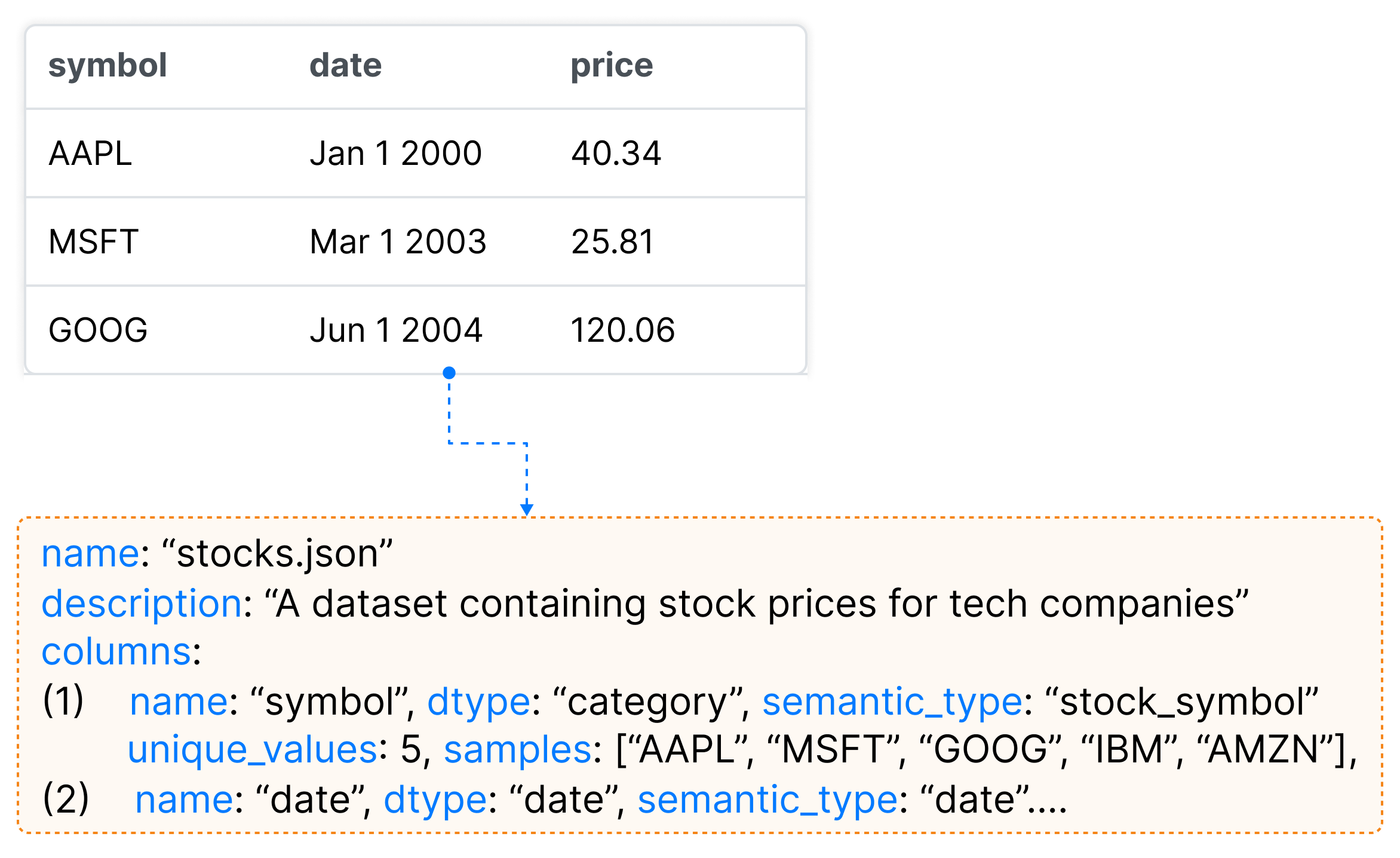}
\caption{The Summarizer constructs an NL summary from extracted data properties.}
\label{fig:summary}
\end{minipage}
\hfill
\begin{minipage}[b][][b]{0.55\linewidth}
\centering
\includegraphics[width=0.9\textwidth]{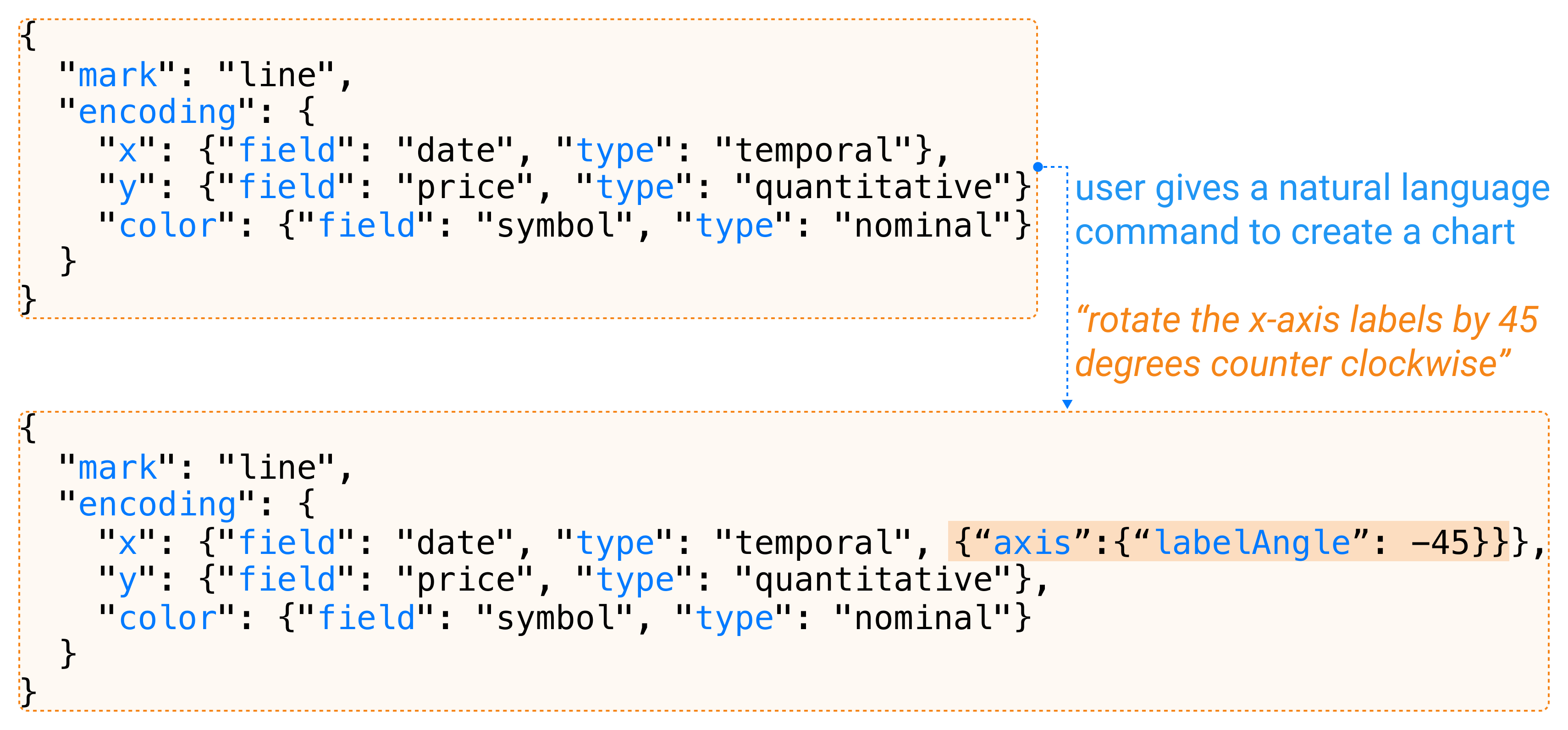}
\caption{Given the data summary, the NL prompt, and an (optional) chart specification, the chart engine produces an updated chart specification.}
\label{fig:vissynth}
\end{minipage}%
\end{figure}

\subsubsection{Pre-processing: Data Summary and Visualization Pre-processing}\hfill

To generate visualizations and synthesize dynamic widgets, the LLM needs an accurate context of the data the user is working with. However, due to the limited context window supported by LLMs, we cannot pass the whole data to the LLM. Hence, to augment LLMs with grounding context about the data, we borrow the \textsf{Data Summarizer} from Lida~\cite{dibia2023lida}. This summarizer is used to produce a dense yet compact summary for any given dataset that is useful as a grounding context for visualization tasks. For every LLM query in the subsequent steps, we pass the data summary instead of the data. First, the summarizer applies rules to extract dataset properties including atomic types (e.g., integer, string, boolean), general statistics (min, max, unique values, etc.), and a random list of n samples for each column using the pandas python library~\cite{pandas}. Then the base summary is enriched by an LLM to include a semantic description of the dataset (e.g., a dataset of stock prices for top 5 tech companies for 10 years), and fields (e.g., Stock price in USD) as well as field semantic type prediction~\cite{zhang2019sato} (see Figure~\ref{fig:summary}).

Before sending any edit or widget synthesis queries to the LLM, \tool splits the data from the visualization specification, and any other unnecessary chart properties (like configuration properties like width, height, etc.) to prevent context overflow. This also ensures that the rendering engine can remain responsive to UI changes. 
\subsubsection{LLM-based Synthesis of Visualization and Dynamic Widgets}\hfill

\paragraph{\bf{Synthesizing visualization.}}
The \textit{Chart Engine} is primarily responsible for synthesizing visualizations --- either new visualizations or editing existing visualizations based on the natural language prompt from the user. The Chart Engine uses the enriched data summary from the summarizer, the user prompt, and optionally an existing \vgl specification to return a \vgl specification using a LLM [Fig~\ref{fig:vissynth}]. To ensure that the LLM produces a valid \vgl specification, and to prevent version-based errors, we instruct the model to only use the \vgl \textsf{Schema v5} and to produce a valid JSON output in a markdown code-block format with language descriptors. Since chat-based models like GPT 3.5 are trained to produce verbose output, using a Markdown code-block format ensures we can reliably extract the code block. To maintain consistency of output, we provide some few-shot learning examples of \vgl specification with its description to the model. 

\paragraph{\bf{Synthesizing dynamic widgets.}}
The \textit{Widget Engine} is primarily responsible for synthesizing dynamic widgets. The widget engine uses the data summary, current \vgl chart specification, and the user prompt to return the widget components--both HTML script and the Javascript callback function. 

To ensure that the LLM produces a valid program for the widget we provide strict templates for HTML and JavaScript. The HTML template defines the empty \textsf{<div>} stub with commented instructions. The JavaScript template has an empty JS function stub with a predefined return value. This template improves the reliability and predictability of the code generated. Templates also help us verify the code in the next step of Program Analysis. We also provided a few-shot examples of HTML and JS code to the model to maintain consistency of output.

\subsubsection{Post-Processing through program analysis}\hfill

\paragraph{\bf Processing visualizations} Once we extract the JSON specification from the markdown output, we parse the specification to check for JSON formatting errors and then compile the \vgl specification to check for syntax or schema errors in the synthesized specification. In case of errors, we provide the error message and ask the model to fix the errors in the same conversation context. If this doesn't work, we re-try the prompt once again.

\paragraph{\bf Processing dynamic widgets} To prevent errors, and ensure the validity of the widgets, \tool performs a series of post-processing steps on the HTML and JS code synthesized by the LLM using program analysis. We mention some of the many steps involved in the post-processing stage below. Each step is achieved by parsing the HTML and JS code to an abstract syntax tree (AST) and manipulating the AST.
\begin{itemize}
    \item Parse the HTML code, to ensure there are no conflicts in the HTML \say{\textsf{ID}} property between the synthesized widgets and previous widgets. If we find conflicts, we programmatically modify the ID property and modify the corresponding JS callback function.
    \item Parse the synthesized JS callback function to ensure it has the right function name and valid function parameters.
    \item Identify and replace the HTML IDs used in the callback function that were modified in the HTML script.
    \item \tool ensures that the properties of the chart being edited are either already present in the current chart or the callback function handles the \textsf{null} case correctly. 
\end{itemize}

\subsection{User Interface Implementation}
\tool is implemented as a web application with React and Typescript. We use the \textsf{html-react-parser}~\cite{npm-html-react-parser} library to attach and detach widgets on the fly as they are created and deleted. \tool's backed hosts the data summarizer (implemented in pandas) implemented as a flask web server that communicates with the front-end using REST API. We use the OpenAI API~\cite{openai-api} to issue queries to the LLM. We chose \textsf{gpt-3.5-turbo} as the target LLM in our current implementation from OpenAI because it strikes the right balance of accuracy vs. speed. Since we had to make edits and synthesize widgets within interactive time to prevent annoying the users, the GPT-3.5 model was responsive enough and accurate enough to suit our needs. 
We have also tested our tool with the more advanced GPT-4 model, which supports longer context and has better instruction-following capability to generate widgets more accurately. However, the latency is too high for smooth interaction. 

\section{User Study Design}

To understand how users can use \tool to solve visualization editing tasks, we conducted a within-subjects lab study with 24 participants. In the study, users are asked to solve two sets of five visualization editing tasks, one set using \tool and another using a baseline NLI-based tool. We aim to answer the following research questions:

\begin{itemize}
    \item [\textbf{RQ1}] Does \tool reduce users' efforts to edit visualizations?
    \item [\textbf{RQ2}] In what scenarios do users prefer to use dynamic widgets compared to a baseline tool?
    \item [\textbf{RQ3}] What are users' strategies to work with dynamic widgets? 
\end{itemize}

\subsection{Participants}
We recruited 24 participants (13 female, 10 male, 1 chose not to disclose) through the mailing lists of two research universities. Of the 24 participants, 2 worked with data visualization at least once daily, 4 participants worked with visualizations weekly, 14 participants worked with visualizations at least once a month, and 4 participants less frequently but still occasionally worked with data visualization. Participants mentioned they had prior experience with a variety of visualization tools and libraries including matplotlib (Python), Seaborn (Python), ggplot (R), Tableau, Excel, D3, and more. None of the participants had any experience with \vgl or Vega libraries. Three of the 24 participants reported they performed data analysis daily, 6 participants did it weekly, 8 participants did it at least a few times a month, and 7 participants did it occasionally (less than a few times a month). Participants received a \$25 Amazon gift card as compensation for their time.

\subsection{Study Conditions} We consider the following two conditions in our user study. We choose the NL-based visualization editing approach as the baseline~\cite{dibia2019data2vis}.

\begin{itemize}
    \item \textbf{Baseline Condition (\control): } We modified the \tool interface with the support for Natural Language Commands along with a set of pre-populated UI widgets for basic chart manipulation (e.g., chart title, axis range, etc.) and remove the support for dynamic widgets. By providing both NL and basic UI supports for users to freely choose from, we believe this is a fair state-of-the-art baseline tool for the study. Pre-populated widgets are similar to static UI users use to edit visualizations (like Excel, Google Sheets, etc.). For the editing tasks, users can use a combination of natural language commands and the pre-populated widgets to edit the chart.

    \item \textbf{Experiment Condition (\experiment):} This is \tool with the support for dynamic widgets on top of \control. With this UI, the users can edit the chart using natural language commands, add custom dynamic widgets, or use the same set of pre-populated widgets. As an addition, whenever the user provides a natural language edit command to the AI, we will synthesize and automatically add a dynamic widget. We display the synthesized widgets in reverse chronological order, so the latest synthesized widget is shown at the top.
\end{itemize}

Note that we didn't explicitly set up a widget-only baseline based on an existing visualization tool as it would require us to restrict participants to only people with experience with a certain visualization tool, hence limiting the diversity of the participants. However, we do collect user feedback in the interview on their opinions on how their experiences with \tool differ from their favorite tools.

\subsection{Tasks}

We selected three data visualization tasks derived from popular datasets. Since our focus is on visualization editing tasks, not the authoring task, the participant starts each task with a base dataset and a base visualization. Informed by the formative study conducted by~\citet{wang2022towards}, the participants will have to perform a series of edits to the visualization. Each task contains five sub-tasks containing instructions to perform edits to the base visualization. We include the following three types of questions: (1) editing the visualization with a concrete editing task, (2) exploratory editing task (e.g., try a few options and then pick the best stroke width), and (3) editing tasks with a question to be answered based on the chart (e.g., zoom in to a time window or range, filter data, etc.).

\paragraph{Task 1 (Stock Trends).} Given a dataset of stock prices for the top five tech companies over ten years and the baseline chart below visualizing the stock trend, the user is asked to complete the following sub-tasks.
\noindent\begin{center}
\begin{minipage}[h]{0.32\linewidth}
\includegraphics[width=\linewidth]{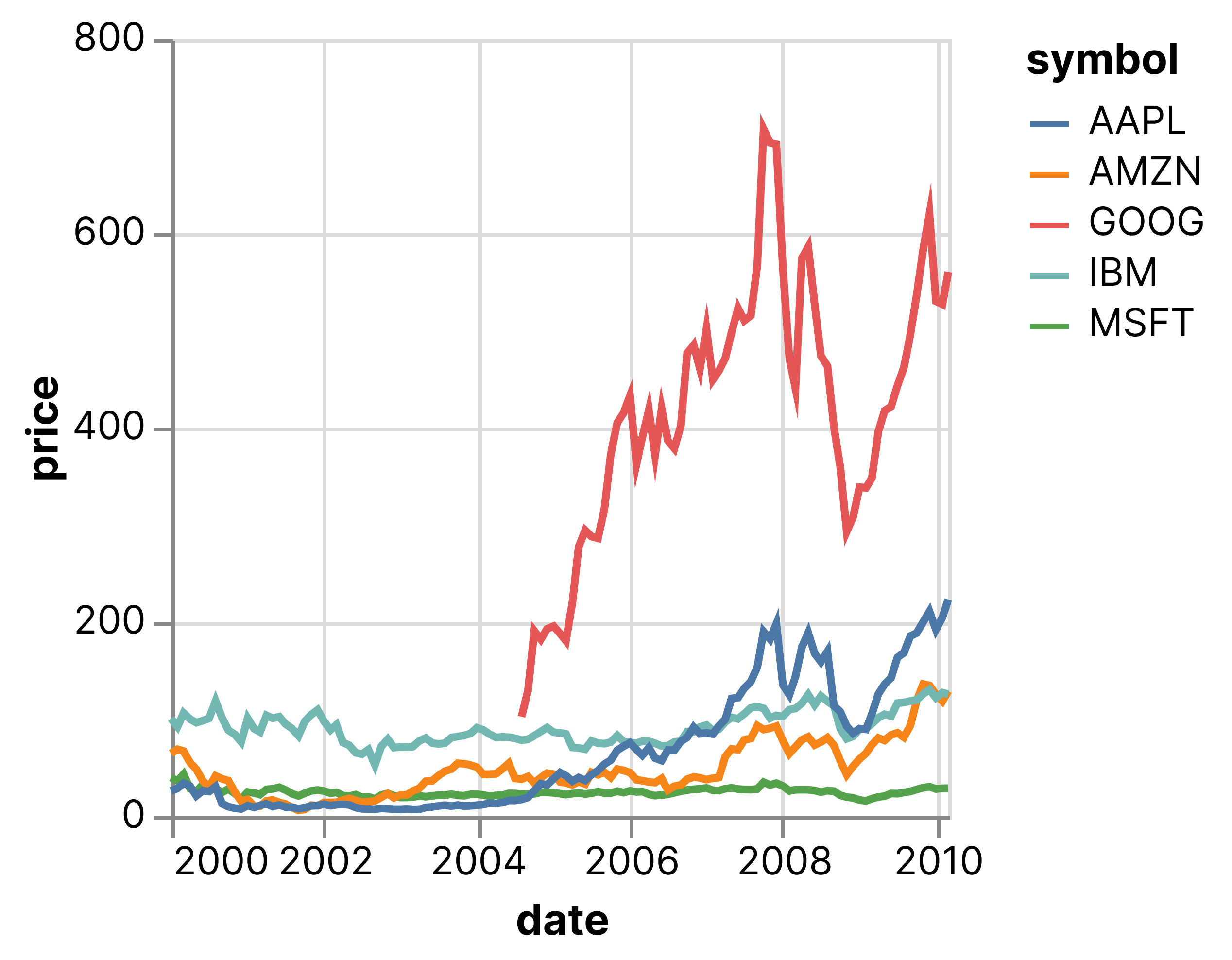}
\end{minipage}
\begin{minipage}[h]{0.67\linewidth}
{\small \begin{enumerate}
    \item Change the chart title to ``Stock Trend''.
    \item Try different lines stroke widths and find the best one that suits you.
    \item Edit the chart to show the stock trends for AAPL and MSFT only.
    \item Change the y-axis max range to 250 to get a zoomed in view. Then, answer the question: ``By just viewing the chart, get the minimum and maximum stock price for both AAPL and MSFT''.
    \item Now compare only between IBM and MSFT. Then, answer the question: ``By just viewing the chart, find the month and year, where the difference in stock price between the two companies were maximum''.
\end{enumerate}}
\end{minipage}
\end{center}

\paragraph{Task 2 (Unemployment Data)} Given a dataset of USA unemployment statistics over multiple sectors and the stacked area chart below visualizing the distribution of unemployment numbers, the user is asked to complete the following sub-tasks.
\noindent\begin{center}
\begin{minipage}[h]{0.32\linewidth}
\includegraphics[width=\linewidth]{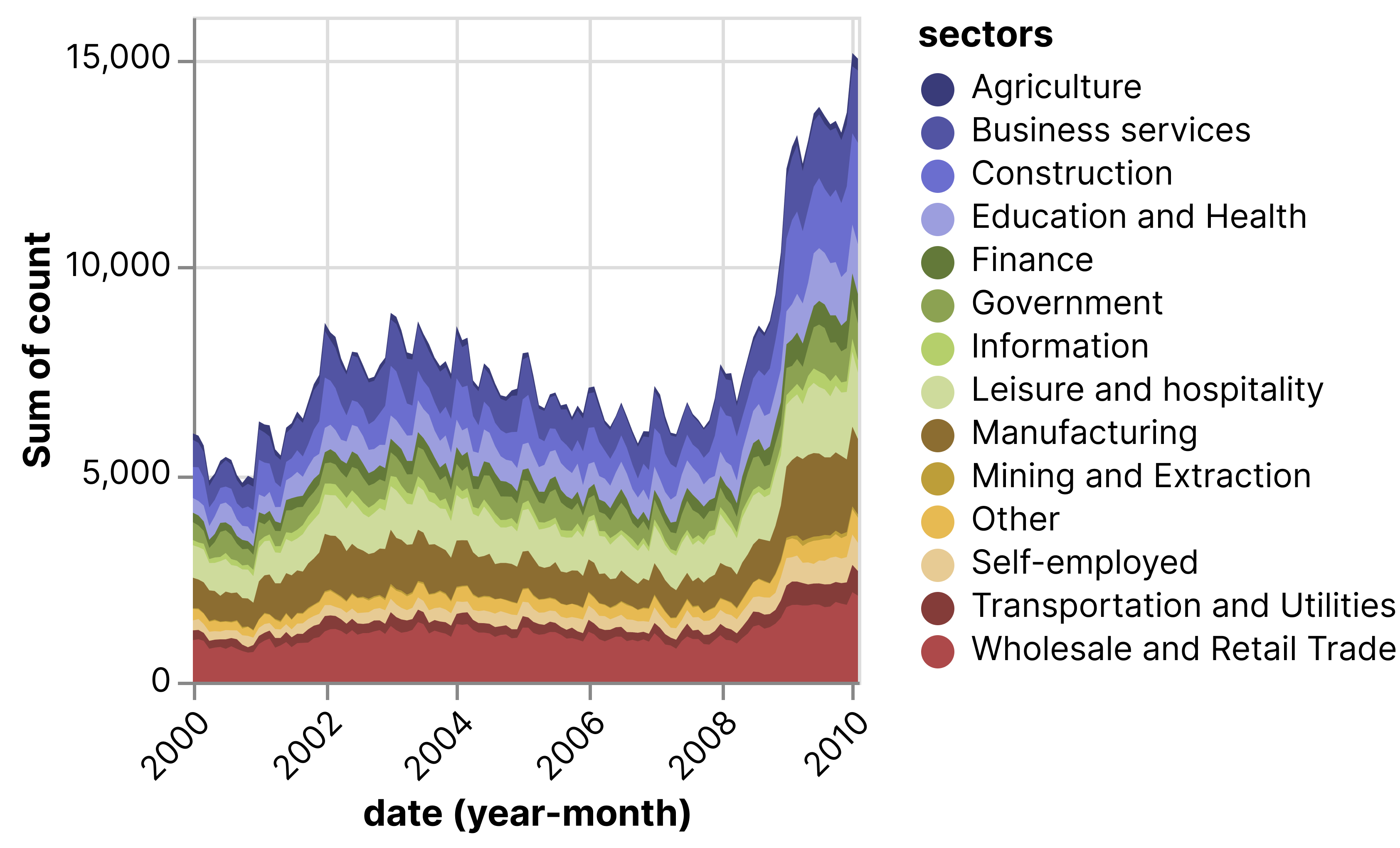}
\end{minipage}
\begin{minipage}[h]{0.67\linewidth}
{\small \begin{enumerate}
    \item Change the x-axis title to ``Timeline''.
    \item Try out different legend positions (inside and outside the chart) to choose the best position that suits you.
    \item Edit the chart to show the trends for Construction and Agriculture sectors only.
    \item Edit the y-axis max date to 06/01/2004 (June 2004). Then, answer the question: ``By just viewing the chart, get the approximate month and year where the difference in the unemployment rate between the two sectors were maximum''.
    \item Now compare only between Finance and Construction. Then, answer the question: ``By just viewing the chart, get the approximate month and year where the difference in the unemployment rate between the two sectors were maximum''.
\end{enumerate}}
\end{minipage}
\end{center}

\paragraph{Task 3 (Weather Data).} Given a dataset of Seattle weather for ten years and the stacked bar chart below visualizing the aggregated distribution of weather over each month, the user is asked to complete the following sub-tasks:
\noindent\begin{center}
\begin{minipage}[h]{0.32\linewidth}
\includegraphics[width=\linewidth]{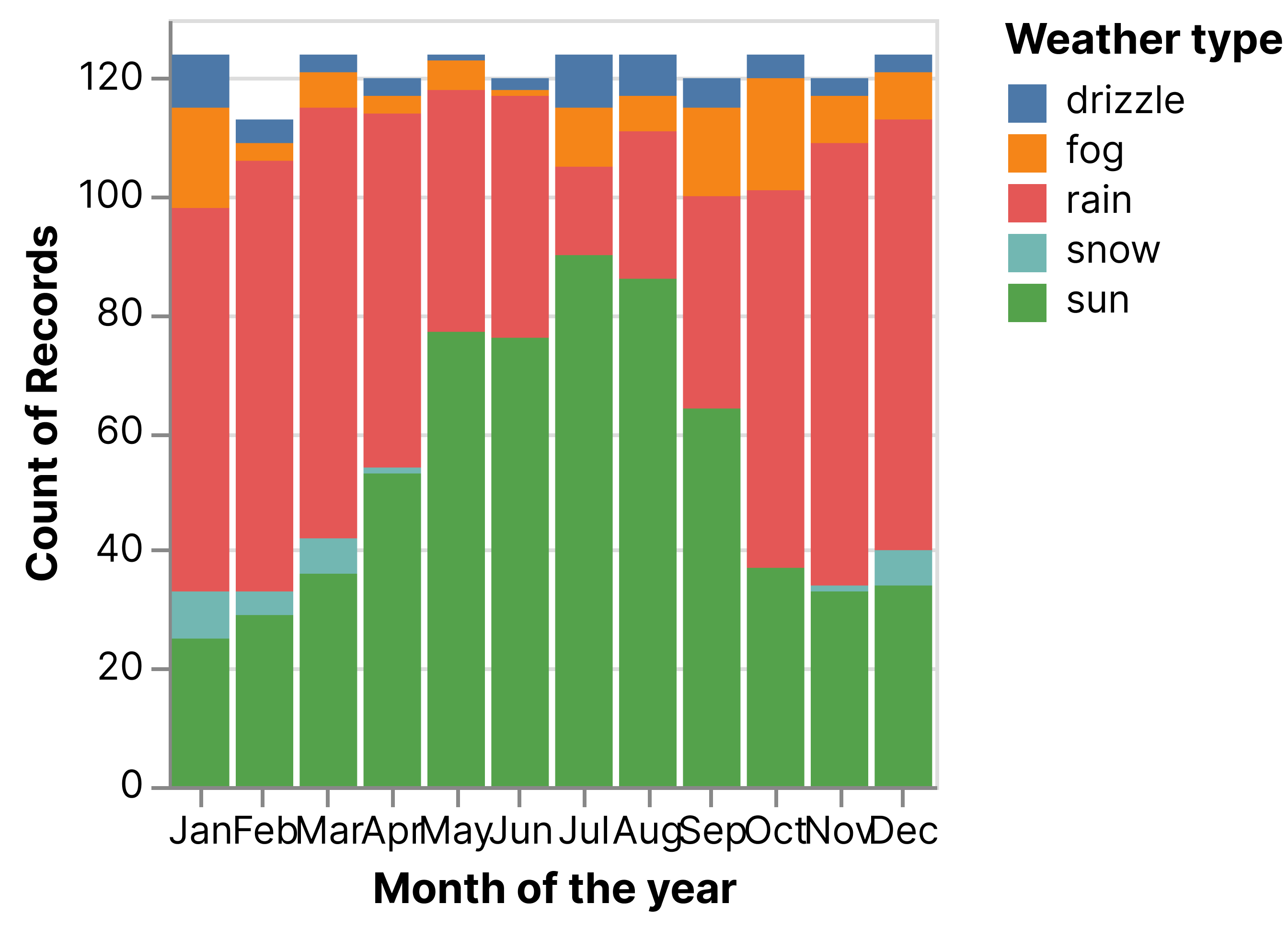}
\end{minipage}
\begin{minipage}[h]{0.67\linewidth}
{\small    \begin{enumerate}
        \item Change the y-axis title to ``Number of Records''.
        \item Rotate the x-axis labels by 45, 65, and 90 degrees and choose the best one for you.
        \item Change the color of each weather type to the following colors: Sun to Yellow, Snow to Gray, Rain to Blue, Fog to Green, Drizzle to Purple. 
        \item Edit the chart to only show ``Snow'' weather for all the months. Then, answer the question: ``By just viewing the chart, find the month with the lowest snow days''.
        \item Edit the chart to only show ``Fog'' weather for all the months. Then, answer the question: ``By just viewing the chart, find the months with the highest fog days''.
    \end{enumerate}
}
\end{minipage}
\end{center}

\subsection{Study Procedure}
To enable easy access to \tool, we hosted both the control and the experiment versions of the tool online which could be accessed by participants via their web browser. After obtaining user consent, we recorded the audio and the screen-cast of each participant, and the users were encouraged to think aloud during the study. In each study session, the participant completed one of the three tasks using the control condition and another task with the experiment condition. To mitigate the learning effect, both the order of task assignment and the order of tool assignment were counterbalanced across participants through random assignment. Therefore, for each unique combination of 3 tasks and 2 conditions, we have 8 participant data points. Before each task in a study session, the participant was given a tutorial of the assigned tool and was allowed to explore using the tool for 5 minutes with a test dataset. Before starting each task, we also explained the dataset and the base visualization and provided time for the participants to explore and understand the dataset. We set a time limit of 15 minutes for each task. After each task, the participant filled out a post-task survey to reflect on their experience using the tool. After finishing both tasks, participants answered a final survey to directly compare the two conditions. At the end of each study session, we also conducted a brief informal interview at the end of the study to get the participant's subjective experience participating in the study and feedback for the tool. 

\subsection{Measurements and Analysis}

We recorded both quantitative and qualitative metrics during the user study. We measured the success/failure for each editing sub-task the participant had to perform. A sub-task is considered failed if the participant is unable to finish the task despite multiple attempts with the tool. They are allowed to retry as many times as they prefer. Through app telemetry, we also recorded all the natural language commands the user provided to the tool and the interactions with the widgets in the tool. In the post-task survey the user filled after every task, we recorded self-reported NASA Task Load Index, self-reported Likert scores for ease of completing the task, and how well the AI understood their intent (for survey questions look at Table~\ref{tab:posttask}). In the post-study survey the user filled at the end of both the tasks, we recorded the participant's self-reported preference and modified the NASA Task Load Index that focused directly on comparing their experience between the two tools(for survey questions look at Table~\ref{tab:poststudy}). For qualitative analysis, the first author performed open-coding on the participants' responses, and the audio transcripts to identify themes, and then discussed with co-authors to refine the themes over multiple sessions. These themes are used to explain the qualitative results. We use paired t-test to measure the statistical significance of quantitative metrics.

\begin{table*}\small
\begin{tabular}{l} \hline
Q1.1. It was easy to complete the tasks using the tool provided. (1-Strongly Disagree, 7 - Strongly Agree)\\
Q1.2. The AI understood my intent and made the right edits. (1-Strongly Disagree, 7-Strongly Agree)\\
Q2.1. How mentally demanding was this task with this tool? (1—Very Low, 7—Very High)\\
Q2.2. How hurried or rushed were you during this task? (1—Very Low, 7—Very High)\\
Q2.3. How successful would you rate yourself in accomplishing this task? (1—Perfect, 7—Failure)\\
Q2.4. How hard did you have to work to accomplish your level of performance? (1—Very Low, 7—Very High)\\
Q2.5. How insecure, discouraged, irritated, stressed, and annoyed were you? (1—Very Low, 7—Very High)\\\hline
\end{tabular}
\vspace{3pt}
\caption{After each task, participants rated (on a 7-point Likert scale) their experience (questions 1.1 - 1.2) and the subjective workload using NASA TLX measures (questions 2.1 - 2.5). }
\label{tab:posttask}
\vspace{-.3cm}
\end{table*}

\begin{table*}
\small
\begin{tabular}{l} 
\hline
Q1.1. Which tool would you prefer to use? (1-\experiment, 7-\control)\\
Q2.1. Which tool was more mentally demanding to communicate? (1-\experiment, 7-\control)\\
Q2.2. Which tool made you feel hurried or rushed during the task? (1-\experiment, 7-\control)\\
Q2.3. Which tool made you feel successful in accomplishing the task? (1-\experiment, 7-\control)\\
Q2.4. For which tool did you work harder to accomplish your level of performance? (1-\experiment, 7-\control)\\
Q2.5. Which tool made you feel more insecure, discouraged, irritated, stressed, and annoyed? (1-\experiment, 7-\control)\\\hline
\end{tabular}
\vspace{3pt}
\caption{After finishing both the tasks, participants comparatively rated (or a 7-point Likert scale) their tool preference (question 1.1) and the subjective workload using NASA TLX (questions 2.1 - 2.5) comparing between the two conditions. (note: in the survey, the names of the UI were coded to prevent bias )}
\label{tab:poststudy}
\vspace{-.3cm}
\end{table*}
\section{User Study Results}

\subsection{Task Completion}

Participants using the \control tool failed to complete sub-tasks \textit{more} often than the participants using the \experiment tool. When using the \control tool, seven participants (P4, P5, P6, P10, P11, P16, P17) failed one of the sub-tasks compared to two participants (P1, P20) when using the \experiment tool. None of the participants failed on more than one sub-task during the study. It is important to note, that these failures are despite participants retrying the tasks as many times as they want. Figure~\ref{fig:res-tlx-pre} shows the self-reported score (Likert scale; higher is better) for ease of completing the task with the tool provided for each condition. Participant using \experiment found it significantly ($p = 0.004$) easier to complete the task ($\mu = 6.26$, $\sigma=0.86$) compared to using the \control ($\mu = 5.13$, $\sigma=1.6$).

The average time $\bar{t}$ for participants in the \control to complete the task is \emph{7 minutes and 22 seconds} ($\bar{t}_\text{task\_1}=6'05''$, $\bar{t}_\text{task\_2}=9'23''$, and $\bar{t}_\text{task\_3}=6'37''$). In the \experiment condition, participants took an average time of \emph{6 minutes and 36 seconds} ($\bar{t}_\text{task\_1}=6'20''$, $\bar{t}_\text{task\_2}=7'07''$, and $\bar{t}_\text{task\_3}=6'23''$) to complete the task. The difference in task completion time between the two conditions is not statistically significant. Note that the exploratory nature of some sub-tasks (e.g., task 1.3) encourages participants to spend time exploring the visualization, and the completion speed is not a definitive measure of performance. We report the completion time here as a setup for understanding user efforts that will be discussed in the following sections.

We analyzed the session recordings to identify the root cause of these task failures. Six of the seven failures in the \control condition and both the failures in the \experiment condition happened in Task 2.4, where the participants had to slice the date range for the chart. This was due to the model returning the \textit{dates} in an incorrect format when editing the \vgl spec from an NL command. \vgl uses a specific JSON date format, e.g. \texttt{\{\say{date}: 14, \say{month}: 3, \say{year}: 2004\}}, whereas the model in many instances used strings to represent the date (e.g. \texttt{\say{2004-03-14}}), resulting in an error. When generating the widget, the model seems to make this mistake fewer times compared to editing the chart specification directly. The remaining failure in the \control condition occurred during Task 1.5, where the participant had to change the filtering values from [MSFT, AAPL] to [MSFT, IBM]. In this instance, the model added an extra conflicting filter transformation instead of editing the existing transformation.

\subsection{Self-reported Cognitive Task Load Index }

\begin{figure}[t]
\centering
\begin{minipage}[b][][b]{0.55\linewidth}
 \includegraphics[width=\linewidth]{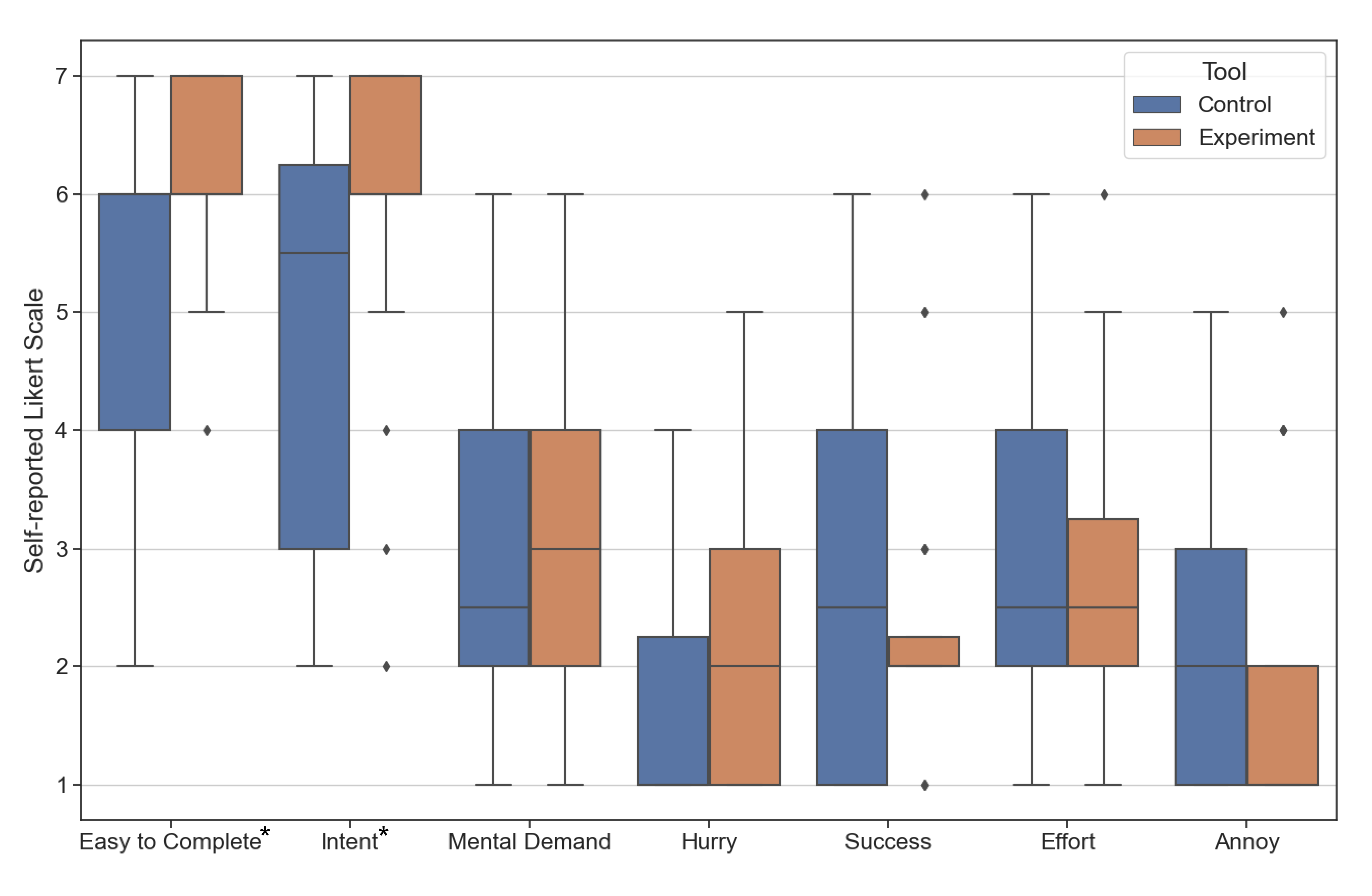}
    \caption{Participants self-reported scores for NASA TLX questions, and ease of completing the task, and how well the AI understood their intent.}
    \label{fig:res-tlx-pre}
\end{minipage}
\hfill
\begin{minipage}[b][][b]{0.43\linewidth}
\centering
    \includegraphics[width=\linewidth]{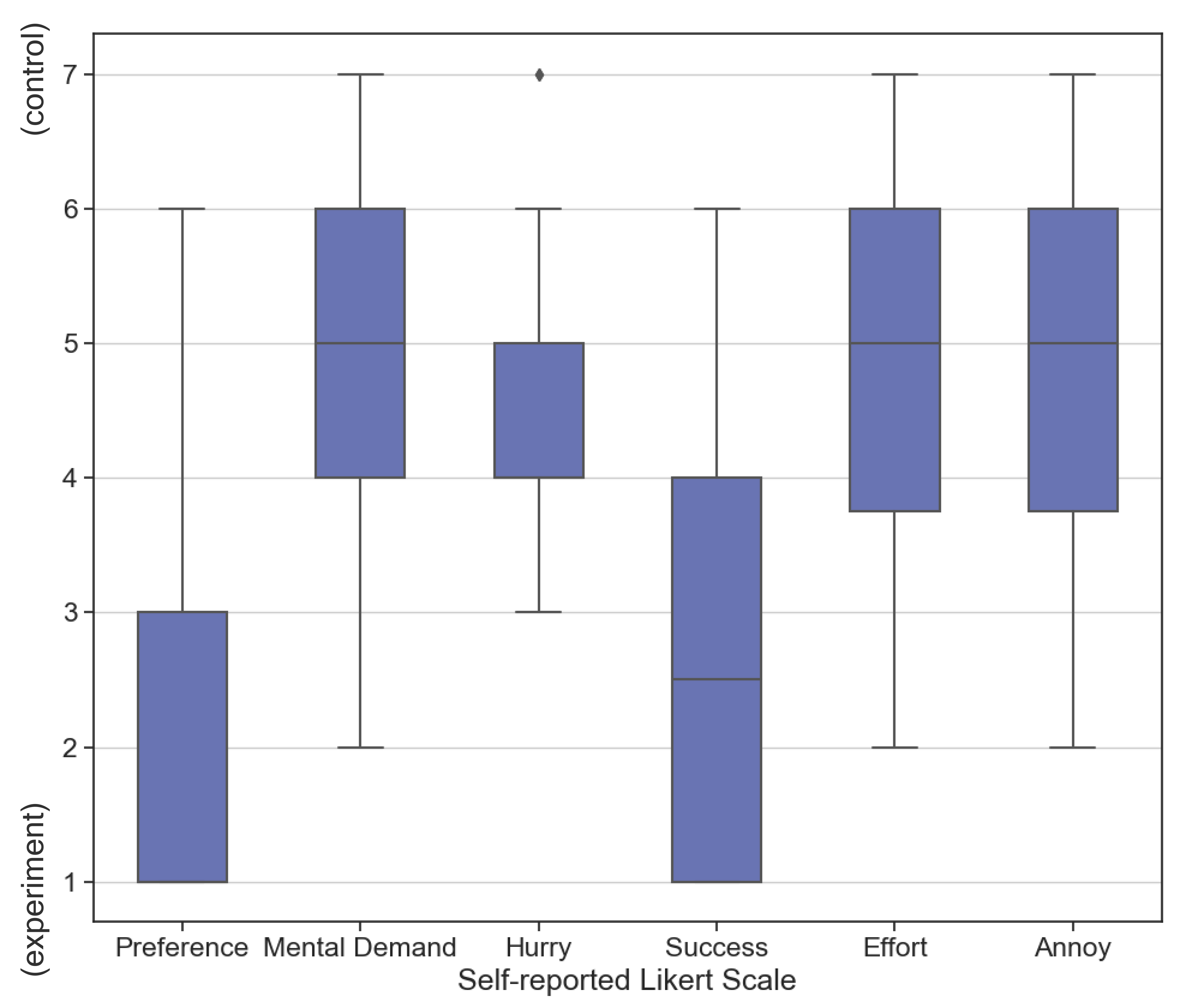}
    \caption{Participants self-reported cognitive load and preference scores that directly compare the two conditions.}
    \label{fig:res-tlx-post}
\end{minipage}%
\end{figure}

In the post-task self-reported NASA TLX ratings where participants scored their cognitive load performing the tasks in both the conditions (check Table~\ref{tab:posttask} for questions), we did not find any statistically significant difference in the mental demand, how hurried or frustrated they felt, the effort required to complete the tasks and their perception of success (see Figure~\ref{fig:res-tlx-pre}).

In the post-task survey where the participants directly compared their cognitive load between the two conditions (see Table~\ref{tab:poststudy} for questions), when using the \experiment tool, 80\% of participants felt less mental demand, 83\% of participants felt less hurried, 62\% of participants felt more successful, 75\% of participants spent less effort, and 75\% of the participants felt less frustrated compared to when using \control tool (see Figure~\ref{fig:res-tlx-post}).

\subsection{User Behavior}

\subsubsection{Natural Language Commands vs. Dynamic Widgets usage}\label{sssec:usage}

\begin{figure}[h]
    \centering
    \includegraphics[width=0.5\linewidth]{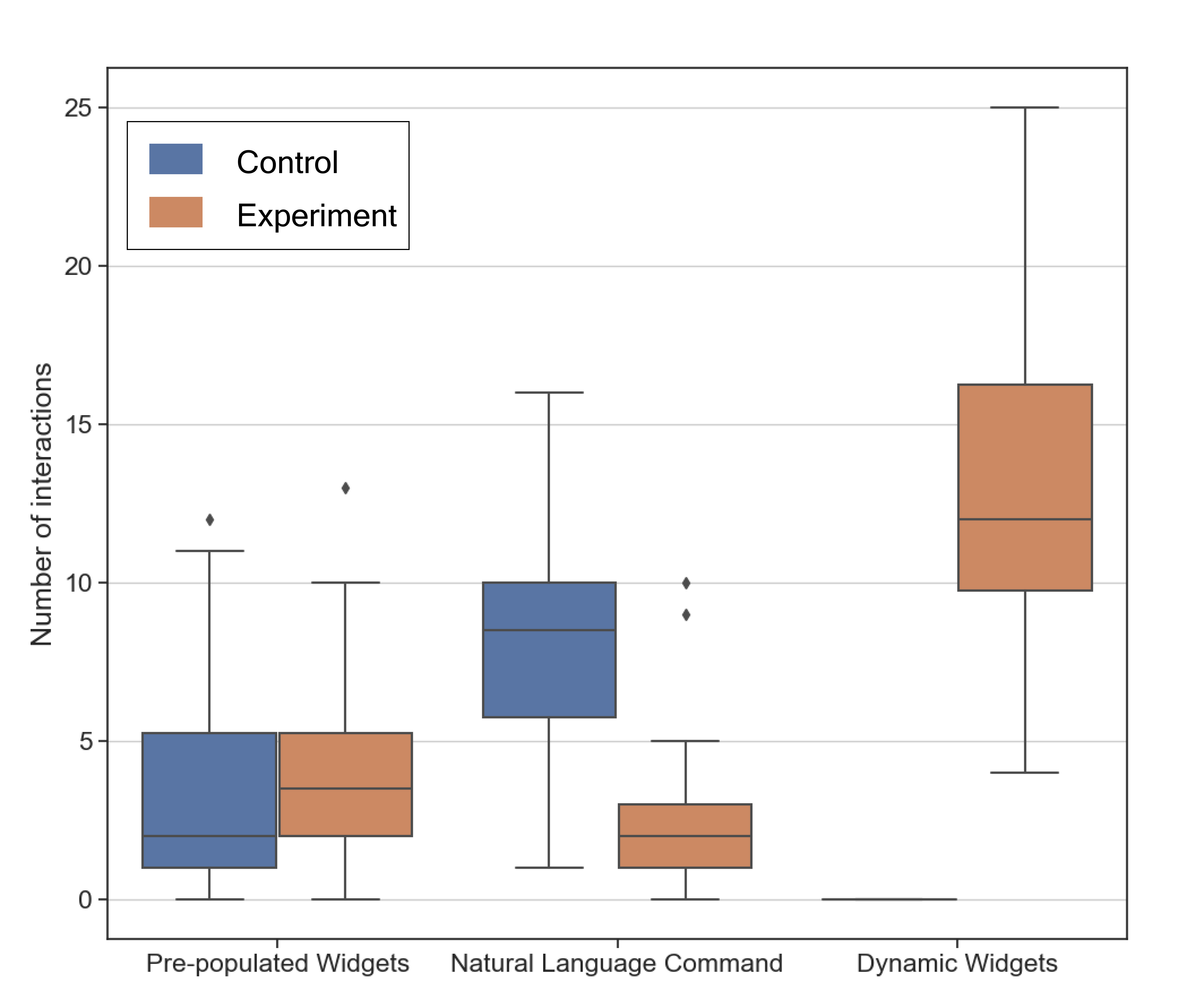}
    \caption{Usage metrics show the number of times the user used pre-populated widgets, dynamic widgets, and natural language commands.}
    \label{fig:telemetry}
\end{figure}

Figure~\ref{fig:telemetry} shows the usage data for the number of natural language commands invoked by the participants, and the number of interactions with both dynamic widgets and pre-populated widgets. 

In the \control condition, participants used an average of 8.4 natural language commands per task (including all sub-tasks) to edit the visualization. This usage significantly ($p < 0.001$) decreased to just 2.67 natural language commands when using the \experiment condition. This was indeed replaced by the dynamic widget usage, where participants on average used it 13.25 times when using the \experiment condition. This confirms our observation of an increase in the number of times participants fine-tuned when using \experiment condition. The widgets reduced the barrier to trying out many different values before settling on the edits. For instance, P6 said \sayit{With dynamic widgets, I like how easy it is toggle or make edits in small increments and try out many different values. I can see what I’m clicking and have fine grained controls.} There was no significant difference in the usage of pre-populated widgets between the two conditions.

\subsubsection{User Strategy}

In the \control condition, we observed two major strategies used by participants to perform the tasks. Six of 24 participants (P7, P9, P12, P13, P15, P17) always used natural language commands to edit the visualization irrespective of any pre-populated widgets available to perform the edit. Whereas, the other 18 participants first looked for a pre-populated widget to perform their edits before resorting to natural language commands.

In the experiment condition \experiment, we observed two major strategies used by participants. Nine participants (P2, P3, P4, P5, P6, P8, P17, P18, P21) preferred to use only the widgets to make all the edits. These participants used the AI only to synthesize dynamic widgets without making the edits and then chose to make the edits by interacting with the synthesized widget instead. Ten participants (P7, P10, P12, P13, P16, P19, P20, P22, P23, P24) preferred to use the natural language command to perform the broad edit, and then use the automatically synthesized dynamic widgets to make fine-grained edits. P23 said \sayit{I would prefer to use the edit prompt for the first time, and use the dynamic widgets for further edits and adjustments}.
P15 chose to use only natural language commands to make the edits and said \sayit{I think that [\control] was easier to use. It is quite easy to make edits quickly using AI commands}. The remaining participants used a combination of widgets and NL commands with no particular preference. 

\subsubsection{Prompting Strategy}

Participants when prompting the AI to edit the chart generally provided their intent (goal) to the AI. For all the sub-tasks, they structured their prompt as \textsf{[ACTION VERB] + [CHART PROPERTY] + [VALUE]}. Some examples are \sayit{Change the stroke width to 5}, \sayit{Rotate the x-axis labels to 65 deg}, \sayit{change the x-axis label to timeline}. 

However, we did notice a variety of natural language commands for filtering-based sub-tasks (Tasks 1.3, 1.5, 2.3, 2.5, 3.4, 3.5). Some participants directly mentioned the values to keep. e.g., \sayit{show me finance and construction sectors only}. Some participants mentioned values to remove. e.g., \sayit{Remove everything other than MSFT and AAPL} or \sayit{Remove the plot for AAPL. Show me the variation of IBM}. Some participants who weren't proficient with charting libraries used incorrect chart property names like \sayit{visibility of the strokes with only MSFT and AAPL selected}, \sayit{remove all legend categories except construction and agriculture}. In three such instances, the model still performed the right filter transformation. For two instances, the model tried to modify the visibility resulting in an incorrect chart. 

When prompting the AI for widgets, participants skip providing the value in the prompt and directly manipulate it via the UI. Their prompt was generally structured as \textsf{[ACTION VERB] + [CHART PROPERTY]}. Some examples are, \sayit{change x-axis limit}, \sayit{change date range}, \sayit{Change the x-axis max date}. For instance, P3 said \sayit{Just an action along with the part of the chart that needed to be edited.}. P23 said \sayit{I keep it as short as possible. Similar to [edit] commands, but without the data.} In the next sub-section, we show evidence of how this makes prompting for a widget easier than using natural language commands to directly edit the chart. By taking advantage of data context, domain constraints, and clever prompt engineering using templates within \tool, we can generate dynamic widgets with very little ambiguity with just simple instructions from users~\cite{zamfirescu2023johnny}. 

\subsection{User Preference}

In the post-task survey, the users were asked to rate their preference between the \control and \experiment conditions. All the participants except P15 (96\%) strongly preferred using \tool compared to the baseline tool. The informal interview and survey responses shed light on the reasons, which we discuss in this sub-section.

\paragraph{\bf Finding 1: Dynamic widgets make repetitive edits easier.}

\noindent Sixteen participants (P1, P3, P4, P5, P6, P7, P9, P11, P13, P14, P17, P18, P19, P21, P22, P23) explicitly mentioned that dynamic widgets greatly reduced effort needed to perform repeated edits. By using dynamic widgets, they reduce the number of natural language commands they need to type -- which is very time-consuming. P5 said,\sayit{I will rather just interact with the dynamic widgets. Compared to typing commands, I would just interact with an UI at the end of the day.}. P7 commented \sayit{Once the widget was there for a certain type of task, I preferred using the widget as clicking is less cumbersome than articulating and typing.}
Moreover, when using dynamic widgets, participants interacted with the LLM only once when creating the widgets. Subsequent edits are performed instantly when compared to using NL commands, where they have to wait for the response from LLM every time they have to perform an edit. P6 said, \sayit{Dynamic widgets were the foundation for the edits I made and I can do however many times I want. Faster than asking the AI again and again.}. This also explains the increased usage of dynamic widgets mentioned in (Section~\ref{sssec:usage}). Users using \experiment, tried a lot more values for exploratory sub-tasks since it was easier to repeat the edits, compared to using \control.

\paragraph{\bf Finding 2: Visual feedback enhances understanding and exploration}

Nine participants (P2, P4, P5, P8, P11, P17, P18, P23, P24) mentioned they prefer using dynamic widgets due to the instant visual feedback the widgets provided when performing the edits. This visual feedback works both ways. First, when the user makes the edits using the dynamic widgets, the edits reflect on the chart immediately, since we aren't waiting for a response from the LLM. P23 summarizes this by saying \sayit{The real-time update with widgets, rather than waiting for AI, was super cool. I can interactively see what happens with the chart.}. It is important to note that these visualizations are inherently static with no interactivity -- but by creating and using dynamic widgets, participants are interacting with an inherently static visualization.

Second, the UI provides visual feedback by retaining the edit information in the state of the UI. P24 explains this by saying \sayit{with dynamic widgets, I can also visualize the changes. For example, if I change the color, I can just see the color there in the widget. That is very helpful, that is just efficient.}. Dynamic widgets provide clear visibility of chart (system) status, improving the visual feedback~\cite{nn-usability}. In other words, dynamic widgets help users overcome the usability challenge of the gulf of evaluation (understanding the system state)~\cite{nn-gulf, norman1986user}. 

\paragraph{\bf Finding 3: Prompting to create dynamic widgets is easier and more reliable}

Figure~\ref{fig:res-tlx-pre} shows the participants' self-reported score (Likert scale; higher is better) for how well the AI understood participants' intent. Participants using \experiment tool ($\mu = 6.04$, $\sigma=1.36$ Likert scale) reported that AI understood their intent significantly better ($p = 0.024$) than participants using \control tool ($\mu = 4.95$, $\sigma=1.77$ Likert scale). 18 of the 24 participants using \control had at least one failed natural language edit command, and 8 of the 24 participants had more than 2 failed natural language edit commands. In contrast, only nine participants using \experiment had at least one erroneous widget, and no participant had more than 2 erroneous widgets when performing the tasks.

Twelve participants (P1, P2, P3, P5, P6, P11, P14, P16, P17, P21, P22, P24) explicitly noted that it was easier and more reliable to prompt for dynamic widgets than to use just natural language to edit the visualization. For instance, P3 commented, \sayit{with [\control] I had to provide my chat command as specific as possible to prevent errors. But with dynamic widgets in [\experiment] I didn’t have to be very specific.}. P6 commented on the reliability by saying \sayit{I like the ability to execute the edits using dynamic widget rather than leave it up to chance using the AI.}

One probable reason for improved reliability could be: when using only natural language commands, the LLM will have to generate the modified \vgl spec for the whole chart in the response. This increases the probability of errors, since during the generation, the LLM can potentially change any other unrelated property in the specification. Whereas, when generating dynamic widgets, the model synthesizes a JavaScript callback function that only edits one (or a few) properties of the \vgl spec, greatly reducing the probability of errors.

\paragraph{\bf Finding 4: Dynamic widgets enhance the sense of control over NL}

Seven participants (P3, P5, P6, P8, P10, P16, P20) mentioned that using dynamic widgets provided them a greater sense of control over the outcome when making edits to the visualization. Users felt that the widgets provided a sense of structure to the edits performed and a sense of autonomy. P8 said \sayit{The widgets gave me a feeling of autonomy to make more changes compared to the AI. It had created some kind of structure for achieving the change.}. Further, P6 said \sayit{typing the text commands was a little non-ideal. Not having the widgets made it feel like I don’t control the edits. [natural language] commands made it feel like there was some level of uncertainty.}

\paragraph{\bf Finding 5: Dynamic widgets enable customization, but can be overwhelming for extended use}

Nine participants (P1, P3, P6, P9, P10, P11, P16, P14, P18) mentioned they liked using dynamic widgets since it allowed them to customize their own UI to suit their editing intents. P7 said \sayit{The dynamic widgets are pretty cool. It’s like writing a whole tool for yourself, but instead of writing code, it is instantly available.} Since we present the widgets in the reverse chronological order of their creation, P18 said \sayit{I like dynamic widgets, because it sort of stores the history of edits I made. I can refer it back whenever I want to change it if I don't like.}.

Compared to traditional static UIs, dynamic interfaces are constantly changing. Five participants (P21, P17, P18, P8, P23) pointed out that creating dynamic widgets can get overwhelming over long editing sessions with widget panels constantly being changed, and wanted some way of either pausing widget creation or ability to categorize and control the widgets that are shown. For instance, P23 said \sayit{ If I made a lot of them [dynamic widgets], then it might get tricky. It’ll be nice to have automatic grouping based on the kind of edits the widgets do.} Similarly, P8 said \sayit{It would be nice to group, collapse, or create multiple relevant and related widgets, just like Photoshop, to customize better.} In contrast, P7 said \sayit{If it was a long series of tasks, I would maybe just prefer a stable interface.}, and P15 said --- \sayit{with [\experiment] I’m not sure how dynamic widgets are helpful. Using the commands was just simpler.}

\subsection{Tool Performance}

To understand the performance of the \tool using the GPT3.5 model, we performed a postmortem analysis to measure the latency faced by the participants due to the model response time, and the number of automatic retries required to generate the correct \vgl specification and widget code.
We measured this by re-playing all the \textit{NL visualization edit} commands, and the \textit{NL add widget} commands provided by the participants during the user study. On average, each query to the GPT3.5 model added a latency of $1.47$ seconds ($\sigma=0.34$ seconds). Also, on average for each NL edit / add widget command, \tool had to automatically retry 1.16 times ($\sigma=0.58$ times) to fix syntax or semantic errors before producing the correct output.
\section{Discussion and Future Work}

\paragraph{\bf Modalities beyond NL and UI widgets} \tool employs two different modalities for users to specify the kinds of edits they want to make to the chart: natural language commands and interaction through dynamic widgets. This design combines the strengths of both modalities so the user can better communicate the intent to the AI agent and quickly make repeated precise edits to the visualization. This was reflected in the study by the participants' preference for the tool, and how they interacted with both modalities. 

An interesting future direction is to expand the modalities beyond the natural language and widget interaction by adding support for voice commands as well as gestures and direct manipulation of the chart. For example, instead of the user giving an NL command \sayit{Move the legend to the bottom of the chart}, the user can simply click and drag (though mouse, touch, or digital pens) the legend to the bottom of the chart to make the edit. The user doesn't need to know they have to modify \textit{legend}, rather they can simply point us toward it. Similarly, \tool also has the potential to include by-example specifications to let users demonstrate editing examples by editing parts of the visualization and then letting the tool generalize edits to other parts of the visualization.

However, multiple modalities also come with their challenges. For example, it is not always easy for users to figure out which modality to use and when. Hence, more research is needed to help users learn the benefits and disadvantages of each modality so that they make an informed choice.

\paragraph{\bf  Static vs. Dynamic UI} One benefit of using natural language commands and dynamic widgets over traditional static UI is the ability to ease the gulf of execution. With static UI, the user has to know how and where to perform the edits, which can be cognitively demanding~\cite{nn-recognition}. Whereas, with NLIs and Dynamic widgets, the users only need to specify their intent. The flip side of this argument is, that despite the learning curve, over time users will learn and get used to the static UIs. However, with dynamic UIs, the interface is constantly changing, which can potentially increase cognitive load for the user, especially in long editing sessions and for users with editing expertise. Some participants from the study did express this concern and suggested having a way of categorizing the widgets predictably. Longer term usability study would provide insights to understand how constantly changing UI affects usability. Unlike traditional interfaces, another limitation of \tool and other purely NL-based interfaces is that they do not present all possible options to the user at all times. This is a double-edged sword; Due to the ad-hoc nature of NL interfaces, NL commands for adding widgets/editing visualizations can sometimes help users discover previously unknown features (similar to observations in~\cite{barke-grounded}), and other times can shift the onus of discovering the tool’s capabilities onto the user leading to mistrust and distrust~\cite{parasuraman1997humans}. More research is needed to study and find methods to overcome these limitations.

Another interesting future direction is to inspect how we can take advantage of dynamic UI widgets' low programming requirement to turn end users into ``no-code developers'' with the ability to customize/DIY their interaction panels to augment static GUI. For example, with dynamic widgets, an end user can construct their own panel that best suits their daily tasks as shortcuts for complex tasks. For example, a user who often works on geographical data analysis can create a custom panel using dynamic widgets specially for map manipulation functions to reduce map editing efforts.

\paragraph{\bf  Supporting imperative plotting libraries and lower-level visualization grammars}
Dynamic widgets are designed around declarative high-level visualization grammar to enable compositional editing (e.g., VegaLite's JSON representation for visualization objects). The declarative syntax helps the widgets to be modular and be synthesized and used in any order the user wishes. Despite their advantages, high-level grammars expose fewer options than low-level grammars or imperative libraries for more complex visualization editing tasks (e.g., to make parts of a line dashed while the rest solid, would require visiting lower-level details of how lines are represented). To support editing of visualization in these low-level languages, we envision combining \tool with bidirectional editing approaches which leverage program analysis and synthesis techniques to propagate surface-level edit requirements to edits over program structures or parameters.

\paragraph{\bf Dynamic widgets for accessibility and other applications}

Prior work on dynamically synthesized UIs stemmed from accessibility research, like SUPPLE~\cite{gajos2004supple} and SUPPLE++~\cite{gajos2007automatically} to accommodate motor and vision capabilities. In the space of dynamic widgets for visualization tasks, there are many possible UIs to perform the same task (e.g., slider vs. number input to control the font size). Every type of UI has accessibility trade-offs based on the user's needs. In the future, we can imagine a version of \tool that lets the user provide their interface constraints and preferences, and \tool automatically synthesizes UI that matches these constraints. Prior work like~\cite{lukes2021synthesis, vaithilingam2019bespoke} has accomplished UI synthesis based on examples or demonstrations. A potential research direction would be to add accessibility constraints and exploit the general knowledge of LLMs to create accessible widgets. 

While \tool is designed for visualization editing tasks, we believe dynamic widgets can also benefit general applications that have extensive configuration options (e.g., document processing software, video processing applications). Users can also take advantage of dynamic widgets' low programming requirements in no-code / low-code tools (like Excel / Tableau) with the ability to customize or DIY their interaction panels to augment existing static UI. Investigating how dynamic widgets can be generalized across different application domains would be worth studying in the future.

\paragraph{\bf \tool design opportunities}

There are a lot of opportunities to improve \tool in the future, some of which we highlight here. In the \control condition, some participants anticipated repeated edits and copied the NL command, for reuse, before submitting it. In the future versions of \tool, we can enable users to access the history of NL commands to make re-running NL commands with small edits easier for the user. One of the most requested features from the participants is for more ways to customize and manage a large number of dynamic widgets. This can involve categorizing widgets by topic (like Adobe Photoshop), and collapsing/expanding (sections of) widgets. Another useful feature is to save and revisit certain combinations of widgets. We can also go further by allowing users to export the visualization spec along with the widgets to be shared with other users, similar to Bespoke~\cite{vaithilingam2019bespoke}.

\section{Related Work}

\paragraph{\bf Visualization authoring tools} Modern visualization authoring tools~\cite{tableau, powerbi-home, ren2018charticulator, satyanarayan2014lyra, liu2018data} and grammars~\cite{satyanarayan2016vega, ggplot2} are built around the grammar of graphics~\cite{wilkinson2012grammar} greatly reduce the visualization authoring efforts by allowing users to specify high-level visualization intent via mapping of data fields to visual properties. For example, users of Tableau or PowerBI can easily drag data fields and drop into encoding shelves of visual properties to specify the mapping, and users of \vgl can provide mappings concisely as a JSON object. Then, based on high-level specification, these tools automatically provide \say{smart defaults} to fill low-level visualization properties (e.g., stroke with, spacing of bars) and compiles the visualization spec to low-level visualization grammars like D3~\cite{d3} for rendering. While such designs reduce the initial visualization authoring complexity, visualization editing, and refinement remain challenging as the user needs to unbox high-level grammar and navigate through the large space of editing options to perform the edits. \tool is designed to address the visualization editing challenge, which complements the strengths of existing authoring systems. We envision that \tool can be combined with existing tools in a way that users start with a high-level specification to describe the visualization intent and then utilize dynamic widgets to perform subsequent edits to refine the chart.

\paragraph{\bf Natural language interfaces for visualization (V-NLIs)}
Natural language interfaces have been extensively adopted to improve the usability of visualization systems~\cite{shen2022towards}. Even commercial GUI-based tools like Tableau~\cite{tableau-nl}, Microsoft Power BI~\cite{powerbi}, and Google Spreadsheets~\cite{dhamdhere2017analyza} automatically translate natural language queries to data queries and present query results with visualizations. However, these systems limit natural language interactions to data queries and corresponding standard charts.

The rapid development of Natural Language Processing (NLP) techniques~\cite{young2018recent, belinkov2019analysis} has provided great opportunities to explore a natural language-based interaction for data visualization. There has been active research in adopting Natural Language Interfaces to improve the usability of visualization systems~\cite{shen2022towards, dhamdhere2017analyza, gao2015datatone, setlur2016eviza, srinivasan2017natural, yu2019flowsense}. With the help of advanced NLP-toolkits~\cite{opennlp, google-nlp, loper2002nltk, spacy, manning2014stanford}, a surge of visualization-oriented Natural Language Interfaces (V-NLIs) have emerged. V-NLI-based authoring systems accept the user's natural language queries or commands as input and output appropriate visualizations. Researchers have explored multiple techniques ranging from heuristics-based approaches to end-to-end learning approaches. 

Heuristic-based approaches explore properties of data in generating a space of potential visualizations~\cite{wongsuphasawat2017voyager}, ranking these space of visualizations based on quality attributes~\cite{luo2018deepeye, moritz2018formalizing} and presenting them to the user. Further works have considered a task decomposition approach, where the user queries are decomposed into multiple tasks. which are then solved individually and then aggregated to yield the final visualization~\cite{narechania2020nl4dv, chen2022type, wang2022towards}. Finally, end-to-end learning-based approaches seek to learn mappings from data directly to generate visualizations~\cite{dibia2019data2vis}. More recently, with the advancements in Large Language Models (LLMs), systems like Lida~\cite{dibia2023lida} have found great success in leveraging patterns learned by LLMs from massive language and code datasets to create visualizations from natural language commands. LLMs preclude the requirement of applying heuristics, or training of custom models paired with custom training and data. As an extension, many V-NLI authoring tools also support visualization editing, with natural language as the primary modality.

\paragraph{\bf Customizable and dynamic user interfaces}
Prior research in domains such as accessibility and ubiquitous computing has worked on systems that automatically generate UIs. SUPPLE~\cite{gajos2004supple} and SUPPLE++~\cite{gajos2007automatically} generate custom UIs for users to accommodate their motor and vision capabilities based on user-provided specifications and activity traces. Projects such as UNIFORM~\cite{nichols2006uniform} and the Personal Universal Controller (PUC)~\cite{nichols2003personal} generate custom UIs for appliances such as media consoles and printers that are customized for each individual's preferences and interaction history. Huddle~\cite{nichols2006huddle} built atop PUC generates UIs to coordinate multiple home electronic appliances. Mavo~\cite{mavo2016} allows users to create interactive HTML pages without the need for programming by just adding special HTML attributes and also provides different editing widgets based on the type of attributes. \tool shares these systems' goals of creating specialized UI tailored to individual users' intent. 

More recent work on Dynamic interfaces follows a \textit{\say{relaxation}} method to create generalized UI widgets. The relaxation method involves creating UI widgets to directly manipulate variables in a function or query. Bespoke~\cite{vaithilingam2019bespoke} synthesizes custom GUIs for command-line applications by using user demonstrations. They employ rule-based heuristics that infer a semantic type for parameters in bash commands to create a dynamic widget for editing the parameter. Similarly, \citet{heer2008generalized} generates dynamic UI using \textit{query relaxing} that enables the users to generalize their selection. A suite of work named precision interfaces~\cite{chen2020monte, chen2022pi2, zhang2018precision} uses SQL queries as a proxy to generate interactive widgets from a sequence of input queries. The latest iteration, NL2Interface~\cite{Chen2022NL2INTERFACEIV} generates SQL queries from NL commands and creates a generalized UI to edit the parameters/variables in the SQL query. BOLT~\cite{srinivasan2023bolt} and EVIZA~\cite{setlur2016eviza} generate ambiguity widgets that provide a simple UI for manipulating values for ambiguous inferred variables. e.g., for the NL command \sayit{largest earthquakes in California}, the threshold for classifying earthquakes as large is ambiguous. However, in both BOLT and EVIZA, natural language commands are restricted by a pre-defined grammar. These tools highlight the importance of complementary GUI tools that accompany NL interfaces. \tool builds on these systems, and uses an LLM to synthesize dynamic widgets that can enable direct manipulation of Visualization properties. Using LLMs to generate dynamic widgets gives us three distinct advantages:

\begin{enumerate}
    \item By providing an accurate representation of the user's context to the LLMs, \tool is less sensitive to errors or ambiguities in natural language commands provided by the user.
    \item We do not have a fixed set of rules, or heuristics, or rely on query relaxation. Instead of synthesizing a UI that allows users to edit just one variable, LLMs can synthesize widgets that can even capture complex relations like manipulating multiple properties at once.
    \item Unlike previous systems, we do not restrict the space or the kinds of UI that can be generated. As LLMs become more powerful, this can enable the synthesis of complex interfaces beyond just the traditional HTML Input elements.
\end{enumerate} 

\paragraph{\bf Multi-modal user interfaces}
Multi-modal interaction techniques have the advantage of letting users better convey their intent in multiple ways reducing the overall effort. Pumice~\cite{li2019pumice} allows users to use natural language to describe programming tasks in end-user development scenarios and then refine intent by providing examples to complement NL's ambiguous nature. DIY Assistant~\cite{fischer2021diy} lets users combine NL and programming specifications to create personal assistants. \citet{lee2019you} enables better sense-making with visual query systems with the help of sketching. ShapeSearch~\cite{siddiqui2020shapesearch} lets users use query shapes using both NL and regular expressions --- greatly improving the expressiveness of shape search queries. Tools like PanaromicData~\cite{zgraggen2014panoramicdata}, and Vizdom~\cite{crotty2015vizdom} allow users to use pen and touch to directly perform data aggregation and analytics respectively on a digital whiteboard. \tool builds on the idea of enabling multiple modalities of interaction. \tool leverages both NL-based interaction to reduce the gulf of execution and UI-based interaction to enhance interactivity. In the future, \tool can further combine pen-and-touch for direct control of visual elements on canvas as well as sketching to demonstrate editing effects.

\section{Conclusion} In this paper, we introduce \tool, which blends natural language and dynamically synthesized UI widgets to ease the gulf of execution and enhance interactivity. Given a visualization edit command or a widget creation command, \tool synthesizes a UI widget that the user can interact with to perform visualization edits. Our study with $24$ participants shows that participants preferred \tool over the NLI-only interface citing ease of further edits and editing confidence due to immediate visual feedback.

\bibliographystyle{ACM-Reference-Format}
\bibliography{99-references}


\begin{thebibliography}{63}


\ifx \showCODEN    \undefined \def \showCODEN     #1{\unskip}     \fi
\ifx \showDOI      \undefined \def \showDOI       #1{#1}\fi
\ifx \showISBNx    \undefined \def \showISBNx     #1{\unskip}     \fi
\ifx \showISBNxiii \undefined \def \showISBNxiii  #1{\unskip}     \fi
\ifx \showISSN     \undefined \def \showISSN      #1{\unskip}     \fi
\ifx \showLCCN     \undefined \def \showLCCN      #1{\unskip}     \fi
\ifx \shownote     \undefined \def \shownote      #1{#1}          \fi
\ifx \showarticletitle \undefined \def \showarticletitle #1{#1}   \fi
\ifx \showURL      \undefined \def \showURL       {\relax}        \fi
\providecommand\bibfield[2]{#2}
\providecommand\bibinfo[2]{#2}
\providecommand\natexlab[1]{#1}
\providecommand\showeprint[2][]{arXiv:#2}

\bibitem[ope(2020)]%
        {openai-api}
 \bibinfo{year}{2020}\natexlab{}.
\newblock \bibinfo{title}{{OpenAI} {API}}.
\newblock \bibinfo{howpublished}{\url{https://openai.com/blog/openai-api}}.
\newblock
\newblock
\shownote{Accessed: 2023-9-13}.


\bibitem[ope(2023)]%
        {opennlp}
 \bibinfo{year}{2023}\natexlab{}.
\newblock \bibinfo{title}{Apache {OpenNLP}}.
\newblock \bibinfo{howpublished}{\url{https://opennlp.apache.org/}}.
\newblock
\newblock
\shownote{Accessed: 2023-9-13}.


\bibitem[goo(2023)]%
        {google-nlp}
 \bibinfo{year}{2023}\natexlab{}.
\newblock \bibinfo{title}{Cloud Natural Language}.
\newblock
  \bibinfo{howpublished}{\url{https://cloud.google.com/natural-language}}.
\newblock
\newblock
\shownote{Accessed: 2023-9-13}.


\bibitem[d3(2023)]%
        {d3}
 \bibinfo{year}{2023}\natexlab{}.
\newblock \bibinfo{title}{{D3} by Observable}.
\newblock \bibinfo{howpublished}{\url{https://d3js.org/}}.
\newblock
\newblock
\shownote{Accessed: 2023-9-14}.


\bibitem[ggp(2023)]%
        {ggplot2}
 \bibinfo{year}{2023}\natexlab{}.
\newblock \bibinfo{title}{ggplot2}.
\newblock \bibinfo{howpublished}{\url{https://ggplot2.tidyverse.org/}}.
\newblock
\newblock
\shownote{Accessed: 2023-9-14}.


\bibitem[pow(2023a)]%
        {powerbi-home}
 \bibinfo{year}{2023}\natexlab{a}.
\newblock \bibinfo{title}{Microsoft PowerBI}.
\newblock \bibinfo{howpublished}{\url{https://powerbi.microsoft.com/en-us/}}.
\newblock
\newblock
\shownote{Accessed: 2023-9-14}.


\bibitem[npm(2023)]%
        {npm-html-react-parser}
 \bibinfo{year}{2023}\natexlab{}.
\newblock \bibinfo{title}{npm: html-react-parser}.
\newblock
  \bibinfo{howpublished}{\url{https://www.npmjs.com/package/html-react-parser}}.
\newblock
\newblock
\shownote{Accessed: 2023-12-6}.


\bibitem[pan(2023)]%
        {pandas}
 \bibinfo{year}{2023}\natexlab{}.
\newblock \bibinfo{title}{pandas documentation --- pandas 2.1.0 documentation}.
\newblock
  \bibinfo{howpublished}{\url{https://pandas.pydata.org/docs/index.html}}.
\newblock
\newblock
\shownote{Accessed: 2023-9-14}.


\bibitem[spa(2023)]%
        {spacy}
 \bibinfo{year}{2023}\natexlab{}.
\newblock \bibinfo{title}{spaCy - Industrial-strength Natural Language
  Processing in Python}.
\newblock \bibinfo{howpublished}{\url{https://spacy.io/}}.
\newblock
\newblock
\shownote{Accessed: 2023-9-13}.


\bibitem[tab(2023)]%
        {tableau}
 \bibinfo{year}{2023}\natexlab{}.
\newblock \bibinfo{title}{Tableau: Business Intelligence and Analytics
  Software}.
\newblock \bibinfo{howpublished}{\url{https://www.tableau.com/}}.
\newblock
\newblock
\shownote{Accessed: 2023-9-14}.


\bibitem[nn-(2023)]%
        {nn-gulf}
 \bibinfo{year}{2023}\natexlab{}.
\newblock \bibinfo{title}{The Two {UX} Gulfs: Evaluation and Execution}.
\newblock
  \bibinfo{howpublished}{\url{https://www.nngroup.com/articles/two-ux-gulfs-evaluation-execution/}}.
\newblock
\newblock
\shownote{Accessed: 2023-9-14}.


\bibitem[pow(2023b)]%
        {powerbi}
 \bibinfo{year}{2023}\natexlab{b}.
\newblock \bibinfo{title}{Use natural language to explore data with Power {BI}
  {Q\&A} - Power {BI}}.
\newblock
  \bibinfo{howpublished}{\url{https://learn.microsoft.com/en-us/power-bi/natural-language/q-and-a-intro}}.
\newblock
\newblock
\shownote{Accessed: 2023-9-13}.


\bibitem[Barke et~al\mbox{.}(2023)]%
        {barke-grounded}
\bibfield{author}{\bibinfo{person}{Shraddha Barke}, \bibinfo{person}{Michael~B.
  James}, {and} \bibinfo{person}{Nadia Polikarpova}.}
  \bibinfo{year}{2023}\natexlab{}.
\newblock \showarticletitle{Grounded Copilot: How Programmers Interact with
  Code-Generating Models}.
\newblock \bibinfo{journal}{\emph{Proc. ACM Program. Lang.}}
  \bibinfo{volume}{7}, \bibinfo{number}{OOPSLA1}, Article
  \bibinfo{articleno}{78} (\bibinfo{date}{apr} \bibinfo{year}{2023}),
  \bibinfo{numpages}{27}~pages.
\newblock
\urldef\tempurl%
\url{https://doi.org/10.1145/3586030}
\showDOI{\tempurl}


\bibitem[Belinkov and Glass(2019)]%
        {belinkov2019analysis}
\bibfield{author}{\bibinfo{person}{Yonatan Belinkov} {and}
  \bibinfo{person}{James Glass}.} \bibinfo{year}{2019}\natexlab{}.
\newblock \showarticletitle{Analysis methods in neural language processing: A
  survey}.
\newblock \bibinfo{journal}{\emph{Transactions of the Association for
  Computational Linguistics}}  \bibinfo{volume}{7} (\bibinfo{year}{2019}),
  \bibinfo{pages}{49--72}.
\newblock


\bibitem[Budiu(2014)]%
        {nn-recognition}
\bibfield{author}{\bibinfo{person}{Raluca Budiu}.}
  \bibinfo{year}{2014}\natexlab{}.
\newblock \bibinfo{title}{Memory Recognition and Recall in User Interfaces}.
\newblock
  \bibinfo{howpublished}{\url{https://www.nngroup.com/articles/recognition-and-recall/}}.
\newblock
\newblock
\shownote{Accessed: 2023-9-12}.


\bibitem[Chen et~al\mbox{.}(2022b)]%
        {chen2022type}
\bibfield{author}{\bibinfo{person}{Qiaochu Chen}, \bibinfo{person}{Shankara
  Pailoor}, \bibinfo{person}{Celeste Barnaby}, \bibinfo{person}{Abby Criswell},
  \bibinfo{person}{Chenglong Wang}, \bibinfo{person}{Greg Durrett}, {and}
  \bibinfo{person}{I{\c{s}}il Dillig}.} \bibinfo{year}{2022}\natexlab{b}.
\newblock \showarticletitle{Type-directed synthesis of visualizations from
  natural language queries}.
\newblock \bibinfo{journal}{\emph{Proceedings of the ACM on Programming
  Languages}} \bibinfo{volume}{6}, \bibinfo{number}{OOPSLA2}
  (\bibinfo{year}{2022}), \bibinfo{pages}{532--559}.
\newblock


\bibitem[Chen(2020)]%
        {chen2020monte}
\bibfield{author}{\bibinfo{person}{Yiru Chen}.}
  \bibinfo{year}{2020}\natexlab{}.
\newblock \showarticletitle{Monte carlo tree search for generating interactive
  data analysis interfaces}. In \bibinfo{booktitle}{\emph{Proceedings of the
  2020 ACM SIGMOD International Conference on Management of Data}}.
  \bibinfo{pages}{2837--2839}.
\newblock


\bibitem[Chen et~al\mbox{.}(2022a)]%
        {Chen2022NL2INTERFACEIV}
\bibfield{author}{\bibinfo{person}{Yiru Chen}, \bibinfo{person}{Ryan Li},
  \bibinfo{person}{Austin Mac}, \bibinfo{person}{Tianbao Xie},
  \bibinfo{person}{Tao Yu}, {and} \bibinfo{person}{Eugene Wu}.}
  \bibinfo{year}{2022}\natexlab{a}.
\newblock \showarticletitle{NL2INTERFACE: Interactive Visualization Interface
  Generation from Natural Language Queries}.
\newblock \bibinfo{journal}{\emph{ArXiv}}  \bibinfo{volume}{abs/2209.08834}
  (\bibinfo{year}{2022}).
\newblock
\urldef\tempurl%
\url{https://api.semanticscholar.org/CorpusID:252367337}
\showURL{%
\tempurl}


\bibitem[Chen and Wu(2022)]%
        {chen2022pi2}
\bibfield{author}{\bibinfo{person}{Yiru Chen} {and} \bibinfo{person}{Eugene
  Wu}.} \bibinfo{year}{2022}\natexlab{}.
\newblock \showarticletitle{Pi2: End-to-end interactive visualization interface
  generation from queries}. In \bibinfo{booktitle}{\emph{Proceedings of the
  2022 International Conference on Management of Data}}.
  \bibinfo{pages}{1711--1725}.
\newblock


\bibitem[Crotty et~al\mbox{.}(2015)]%
        {crotty2015vizdom}
\bibfield{author}{\bibinfo{person}{Andrew Crotty}, \bibinfo{person}{Alex
  Galakatos}, \bibinfo{person}{Emanuel Zgraggen}, \bibinfo{person}{Carsten
  Binnig}, {and} \bibinfo{person}{Tim Kraska}.}
  \bibinfo{year}{2015}\natexlab{}.
\newblock \showarticletitle{Vizdom: interactive analytics through pen and
  touch}.
\newblock \bibinfo{journal}{\emph{Proceedings of the VLDB Endowment}}
  \bibinfo{volume}{8}, \bibinfo{number}{12} (\bibinfo{year}{2015}),
  \bibinfo{pages}{2024--2027}.
\newblock


\bibitem[Dhamdhere et~al\mbox{.}(2017)]%
        {dhamdhere2017analyza}
\bibfield{author}{\bibinfo{person}{Kedar Dhamdhere}, \bibinfo{person}{Kevin~S
  McCurley}, \bibinfo{person}{Ralfi Nahmias}, \bibinfo{person}{Mukund
  Sundararajan}, {and} \bibinfo{person}{Qiqi Yan}.}
  \bibinfo{year}{2017}\natexlab{}.
\newblock \showarticletitle{Analyza: Exploring data with conversation}. In
  \bibinfo{booktitle}{\emph{Proceedings of the 22nd International Conference on
  Intelligent User Interfaces}}. \bibinfo{pages}{493--504}.
\newblock


\bibitem[Dibia(2023)]%
        {dibia2023lida}
\bibfield{author}{\bibinfo{person}{Victor Dibia}.}
  \bibinfo{year}{2023}\natexlab{}.
\newblock \showarticletitle{LIDA: A Tool for Automatic Generation of
  Grammar-Agnostic Visualizations and Infographics using Large Language
  Models}.
\newblock  (\bibinfo{date}{6 March} \bibinfo{year}{2023}).
\newblock
\showeprint[arxiv]{2303.02927}~[cs.AI]


\bibitem[Dibia and Demiralp(2019)]%
        {dibia2019data2vis}
\bibfield{author}{\bibinfo{person}{Victor Dibia} {and}
  \bibinfo{person}{{\c{C}}a{\u{g}}atay Demiralp}.}
  \bibinfo{year}{2019}\natexlab{}.
\newblock \showarticletitle{Data2vis: Automatic generation of data
  visualizations using sequence-to-sequence recurrent neural networks}.
\newblock \bibinfo{journal}{\emph{IEEE computer graphics and applications}}
  \bibinfo{volume}{39}, \bibinfo{number}{5} (\bibinfo{year}{2019}),
  \bibinfo{pages}{33--46}.
\newblock


\bibitem[Fischer et~al\mbox{.}(2021)]%
        {fischer2021diy}
\bibfield{author}{\bibinfo{person}{Michael~H Fischer},
  \bibinfo{person}{Giovanni Campagna}, \bibinfo{person}{Euirim Choi}, {and}
  \bibinfo{person}{Monica~S Lam}.} \bibinfo{year}{2021}\natexlab{}.
\newblock \showarticletitle{DIY assistant: a multi-modal end-user programmable
  virtual assistant}. In \bibinfo{booktitle}{\emph{Proceedings of the 42nd ACM
  SIGPLAN International Conference on Programming Language Design and
  Implementation}}. \bibinfo{pages}{312--327}.
\newblock


\bibitem[Gajos and Weld(2004)]%
        {gajos2004supple}
\bibfield{author}{\bibinfo{person}{Krzysztof Gajos} {and}
  \bibinfo{person}{Daniel~S Weld}.} \bibinfo{year}{2004}\natexlab{}.
\newblock \showarticletitle{SUPPLE: automatically generating user interfaces}.
  In \bibinfo{booktitle}{\emph{Proceedings of the 9th international conference
  on Intelligent user interfaces}}. \bibinfo{pages}{93--100}.
\newblock


\bibitem[Gajos et~al\mbox{.}(2007)]%
        {gajos2007automatically}
\bibfield{author}{\bibinfo{person}{Krzysztof~Z Gajos}, \bibinfo{person}{Jacob~O
  Wobbrock}, {and} \bibinfo{person}{Daniel~S Weld}.}
  \bibinfo{year}{2007}\natexlab{}.
\newblock \showarticletitle{Automatically generating user interfaces adapted to
  users' motor and vision capabilities}. In
  \bibinfo{booktitle}{\emph{Proceedings of the 20th annual ACM symposium on
  User interface software and technology}}. \bibinfo{pages}{231--240}.
\newblock


\bibitem[Gao et~al\mbox{.}(2015)]%
        {gao2015datatone}
\bibfield{author}{\bibinfo{person}{Tong Gao}, \bibinfo{person}{Mira Dontcheva},
  \bibinfo{person}{Eytan Adar}, \bibinfo{person}{Zhicheng Liu}, {and}
  \bibinfo{person}{Karrie~G Karahalios}.} \bibinfo{year}{2015}\natexlab{}.
\newblock \showarticletitle{Datatone: Managing ambiguity in natural language
  interfaces for data visualization}. In \bibinfo{booktitle}{\emph{Proceedings
  of the 28th annual acm symposium on user interface software \& technology}}.
  \bibinfo{pages}{489--500}.
\newblock


\bibitem[Heer et~al\mbox{.}(2008)]%
        {heer2008generalized}
\bibfield{author}{\bibinfo{person}{Jeffrey Heer}, \bibinfo{person}{Maneesh
  Agrawala}, {and} \bibinfo{person}{Wesley Willett}.}
  \bibinfo{year}{2008}\natexlab{}.
\newblock \showarticletitle{Generalized selection via interactive query
  relaxation}. In \bibinfo{booktitle}{\emph{Proceedings of the SIGCHI
  Conference on Human Factors in Computing Systems}}.
  \bibinfo{pages}{959--968}.
\newblock


\bibitem[Lee et~al\mbox{.}(2019)]%
        {lee2019you}
\bibfield{author}{\bibinfo{person}{Doris Jung-Lin Lee}, \bibinfo{person}{John
  Lee}, \bibinfo{person}{Tarique Siddiqui}, \bibinfo{person}{Jaewoo Kim},
  \bibinfo{person}{Karrie Karahalios}, {and} \bibinfo{person}{Aditya
  Parameswaran}.} \bibinfo{year}{2019}\natexlab{}.
\newblock \showarticletitle{You can't always sketch what you want:
  Understanding sensemaking in visual query systems}.
\newblock \bibinfo{journal}{\emph{IEEE transactions on visualization and
  computer graphics}} \bibinfo{volume}{26}, \bibinfo{number}{1}
  (\bibinfo{year}{2019}), \bibinfo{pages}{1267--1277}.
\newblock


\bibitem[Li et~al\mbox{.}(2019)]%
        {li2019pumice}
\bibfield{author}{\bibinfo{person}{Toby Jia-Jun Li}, \bibinfo{person}{Marissa
  Radensky}, \bibinfo{person}{Justin Jia}, \bibinfo{person}{Kirielle
  Singarajah}, \bibinfo{person}{Tom~M Mitchell}, {and} \bibinfo{person}{Brad~A
  Myers}.} \bibinfo{year}{2019}\natexlab{}.
\newblock \showarticletitle{Pumice: A multi-modal agent that learns concepts
  and conditionals from natural language and demonstrations}. In
  \bibinfo{booktitle}{\emph{Proceedings of the 32nd annual ACM symposium on
  user interface software and technology}}. \bibinfo{pages}{577--589}.
\newblock


\bibitem[Liu et~al\mbox{.}(2018)]%
        {liu2018data}
\bibfield{author}{\bibinfo{person}{Zhicheng Liu}, \bibinfo{person}{John
  Thompson}, \bibinfo{person}{Alan Wilson}, \bibinfo{person}{Mira Dontcheva},
  \bibinfo{person}{James Delorey}, \bibinfo{person}{Sam Grigg},
  \bibinfo{person}{Bernard Kerr}, {and} \bibinfo{person}{John Stasko}.}
  \bibinfo{year}{2018}\natexlab{}.
\newblock \showarticletitle{Data illustrator: Augmenting vector design tools
  with lazy data binding for expressive visualization authoring}. In
  \bibinfo{booktitle}{\emph{Proceedings of the 2018 CHI Conference on Human
  Factors in Computing Systems}}. \bibinfo{pages}{1--13}.
\newblock


\bibitem[Loper and Bird(2002)]%
        {loper2002nltk}
\bibfield{author}{\bibinfo{person}{Edward Loper} {and} \bibinfo{person}{Steven
  Bird}.} \bibinfo{year}{2002}\natexlab{}.
\newblock \showarticletitle{Nltk: The natural language toolkit}.
\newblock \bibinfo{journal}{\emph{arXiv preprint cs/0205028}}
  (\bibinfo{year}{2002}).
\newblock


\bibitem[Lukes et~al\mbox{.}(2021)]%
        {lukes2021synthesis}
\bibfield{author}{\bibinfo{person}{Dylan Lukes}, \bibinfo{person}{John
  Sarracino}, \bibinfo{person}{Cora Coleman}, \bibinfo{person}{Hila Peleg},
  \bibinfo{person}{Sorin Lerner}, {and} \bibinfo{person}{Nadia Polikarpova}.}
  \bibinfo{year}{2021}\natexlab{}.
\newblock \showarticletitle{Synthesis of web layouts from examples}. In
  \bibinfo{booktitle}{\emph{Proceedings of the 29th ACM Joint Meeting on
  European Software Engineering Conference and Symposium on the Foundations of
  Software Engineering}}. \bibinfo{pages}{651--663}.
\newblock


\bibitem[Luo et~al\mbox{.}(2018)]%
        {luo2018deepeye}
\bibfield{author}{\bibinfo{person}{Yuyu Luo}, \bibinfo{person}{Xuedi Qin},
  \bibinfo{person}{Nan Tang}, \bibinfo{person}{Guoliang Li}, {and}
  \bibinfo{person}{Xinran Wang}.} \bibinfo{year}{2018}\natexlab{}.
\newblock \showarticletitle{Deepeye: Creating good data visualizations by
  keyword search}. In \bibinfo{booktitle}{\emph{Proceedings of the 2018
  International Conference on Management of Data}}.
  \bibinfo{pages}{1733--1736}.
\newblock


\bibitem[Manning et~al\mbox{.}(2014)]%
        {manning2014stanford}
\bibfield{author}{\bibinfo{person}{Christopher~D Manning},
  \bibinfo{person}{Mihai Surdeanu}, \bibinfo{person}{John Bauer},
  \bibinfo{person}{Jenny~Rose Finkel}, \bibinfo{person}{Steven Bethard}, {and}
  \bibinfo{person}{David McClosky}.} \bibinfo{year}{2014}\natexlab{}.
\newblock \showarticletitle{The Stanford CoreNLP natural language processing
  toolkit}. In \bibinfo{booktitle}{\emph{Proceedings of 52nd annual meeting of
  the association for computational linguistics: system demonstrations}}.
  \bibinfo{pages}{55--60}.
\newblock


\bibitem[Markas({[n.\,d.]})]%
        {tableau-nl}
\bibfield{author}{\bibinfo{person}{Ruhaab Markas}.}
  \bibinfo{year}{[n.\,d.]}\natexlab{}.
\newblock \bibinfo{title}{Ask Data: Simplifying analytics with natural
  language}.
\newblock
  \bibinfo{howpublished}{\url{https://www.tableau.com/blog/ask-data-simplifying-analytics-natural-language-98655}}.
\newblock
\newblock
\shownote{Accessed: 2023-9-13}.


\bibitem[Moritz et~al\mbox{.}(2018)]%
        {moritz2018formalizing}
\bibfield{author}{\bibinfo{person}{Dominik Moritz}, \bibinfo{person}{Chenglong
  Wang}, \bibinfo{person}{Greg~L Nelson}, \bibinfo{person}{Halden Lin},
  \bibinfo{person}{Adam~M Smith}, \bibinfo{person}{Bill Howe}, {and}
  \bibinfo{person}{Jeffrey Heer}.} \bibinfo{year}{2018}\natexlab{}.
\newblock \showarticletitle{Formalizing visualization design knowledge as
  constraints: Actionable and extensible models in draco}.
\newblock \bibinfo{journal}{\emph{IEEE transactions on visualization and
  computer graphics}} \bibinfo{volume}{25}, \bibinfo{number}{1}
  (\bibinfo{year}{2018}), \bibinfo{pages}{438--448}.
\newblock


\bibitem[Narechania et~al\mbox{.}(2020)]%
        {narechania2020nl4dv}
\bibfield{author}{\bibinfo{person}{Arpit Narechania}, \bibinfo{person}{Arjun
  Srinivasan}, {and} \bibinfo{person}{John Stasko}.}
  \bibinfo{year}{2020}\natexlab{}.
\newblock \showarticletitle{NL4DV: A toolkit for generating analytic
  specifications for data visualization from natural language queries}.
\newblock \bibinfo{journal}{\emph{IEEE Transactions on Visualization and
  Computer Graphics}} \bibinfo{volume}{27}, \bibinfo{number}{2}
  (\bibinfo{year}{2020}), \bibinfo{pages}{369--379}.
\newblock


\bibitem[Nichols et~al\mbox{.}(2003)]%
        {nichols2003personal}
\bibfield{author}{\bibinfo{person}{Jeffrey Nichols}, \bibinfo{person}{Brad~A
  Myers}, \bibinfo{person}{Michael Higgins}, \bibinfo{person}{Joseph Hughes},
  \bibinfo{person}{Thomas~K Harris}, \bibinfo{person}{Roni Rosenfeld}, {and}
  \bibinfo{person}{Kevin Litwack}.} \bibinfo{year}{2003}\natexlab{}.
\newblock \showarticletitle{Personal universal controllers: controlling complex
  appliances with GUIs and speech}. In \bibinfo{booktitle}{\emph{CHI'03
  Extended Abstracts on Human Factors in Computing Systems}}.
  \bibinfo{pages}{624--625}.
\newblock


\bibitem[Nichols et~al\mbox{.}(2006a)]%
        {nichols2006uniform}
\bibfield{author}{\bibinfo{person}{Jeffrey Nichols}, \bibinfo{person}{Brad~A
  Myers}, {and} \bibinfo{person}{Brandon Rothrock}.}
  \bibinfo{year}{2006}\natexlab{a}.
\newblock \showarticletitle{UNIFORM: automatically generating consistent remote
  control user interfaces}. In \bibinfo{booktitle}{\emph{Proceedings of the
  SIGCHI conference on Human Factors in computing systems}}.
  \bibinfo{pages}{611--620}.
\newblock


\bibitem[Nichols et~al\mbox{.}(2006b)]%
        {nichols2006huddle}
\bibfield{author}{\bibinfo{person}{Jeffrey Nichols}, \bibinfo{person}{Brandon
  Rothrock}, \bibinfo{person}{Duen~Horng Chau}, {and} \bibinfo{person}{Brad~A
  Myers}.} \bibinfo{year}{2006}\natexlab{b}.
\newblock \showarticletitle{Huddle: automatically generating interfaces for
  systems of multiple connected appliances}. In
  \bibinfo{booktitle}{\emph{Proceedings of the 19th annual ACM symposium on
  User interface software and technology}}. \bibinfo{pages}{279--288}.
\newblock


\bibitem[Nielsen(2020)]%
        {nn-usability}
\bibfield{author}{\bibinfo{person}{Jakob Nielsen}.}
  \bibinfo{year}{2020}\natexlab{}.
\newblock \bibinfo{title}{10 Usability Heuristics for User Interface Design}.
\newblock
  \bibinfo{howpublished}{\url{https://www.nngroup.com/articles/ten-usability-heuristics/}}.
\newblock
\newblock
\shownote{Accessed: 2023-9-13}.


\bibitem[Norman(1986)]%
        {norman1986user}
\bibfield{author}{\bibinfo{person}{Donald Norman}.}
  \bibinfo{year}{1986}\natexlab{}.
\newblock \showarticletitle{User centered system design}.
\newblock \bibinfo{journal}{\emph{New perspectives on human-computer
  interaction}} (\bibinfo{year}{1986}).
\newblock


\bibitem[Parasuraman and Riley(1997)]%
        {parasuraman1997humans}
\bibfield{author}{\bibinfo{person}{Raja Parasuraman} {and}
  \bibinfo{person}{Victor Riley}.} \bibinfo{year}{1997}\natexlab{}.
\newblock \showarticletitle{Humans and automation: Use, misuse, disuse, abuse}.
\newblock \bibinfo{journal}{\emph{Human factors}} \bibinfo{volume}{39},
  \bibinfo{number}{2} (\bibinfo{year}{1997}), \bibinfo{pages}{230--253}.
\newblock


\bibitem[Ren et~al\mbox{.}(2018)]%
        {ren2018charticulator}
\bibfield{author}{\bibinfo{person}{Donghao Ren}, \bibinfo{person}{Bongshin
  Lee}, {and} \bibinfo{person}{Matthew Brehmer}.}
  \bibinfo{year}{2018}\natexlab{}.
\newblock \showarticletitle{Charticulator: Interactive construction of bespoke
  chart layouts}.
\newblock \bibinfo{journal}{\emph{IEEE transactions on visualization and
  computer graphics}} \bibinfo{volume}{25}, \bibinfo{number}{1}
  (\bibinfo{year}{2018}), \bibinfo{pages}{789--799}.
\newblock


\bibitem[Satyanarayan and Heer(2014)]%
        {satyanarayan2014lyra}
\bibfield{author}{\bibinfo{person}{Arvind Satyanarayan} {and}
  \bibinfo{person}{Jeffrey Heer}.} \bibinfo{year}{2014}\natexlab{}.
\newblock \showarticletitle{Lyra: An interactive visualization design
  environment}. In \bibinfo{booktitle}{\emph{Computer graphics forum}},
  Vol.~\bibinfo{volume}{33}. Wiley Online Library, \bibinfo{pages}{351--360}.
\newblock


\bibitem[Satyanarayan et~al\mbox{.}(2016)]%
        {satyanarayan2016vega}
\bibfield{author}{\bibinfo{person}{Arvind Satyanarayan},
  \bibinfo{person}{Dominik Moritz}, \bibinfo{person}{Kanit Wongsuphasawat},
  {and} \bibinfo{person}{Jeffrey Heer}.} \bibinfo{year}{2016}\natexlab{}.
\newblock \showarticletitle{Vega-lite: A grammar of interactive graphics}.
\newblock \bibinfo{journal}{\emph{IEEE transactions on visualization and
  computer graphics}} \bibinfo{volume}{23}, \bibinfo{number}{1}
  (\bibinfo{year}{2016}), \bibinfo{pages}{341--350}.
\newblock


\bibitem[Setlur et~al\mbox{.}(2016)]%
        {setlur2016eviza}
\bibfield{author}{\bibinfo{person}{Vidya Setlur}, \bibinfo{person}{Sarah~E
  Battersby}, \bibinfo{person}{Melanie Tory}, \bibinfo{person}{Rich
  Gossweiler}, {and} \bibinfo{person}{Angel~X Chang}.}
  \bibinfo{year}{2016}\natexlab{}.
\newblock \showarticletitle{Eviza: A natural language interface for visual
  analysis}. In \bibinfo{booktitle}{\emph{Proceedings of the 29th annual
  symposium on user interface software and technology}}.
  \bibinfo{pages}{365--377}.
\newblock


\bibitem[Shen et~al\mbox{.}(2022)]%
        {shen2022towards}
\bibfield{author}{\bibinfo{person}{Leixian Shen}, \bibinfo{person}{Enya Shen},
  \bibinfo{person}{Yuyu Luo}, \bibinfo{person}{Xiaocong Yang},
  \bibinfo{person}{Xuming Hu}, \bibinfo{person}{Xiongshuai Zhang},
  \bibinfo{person}{Zhiwei Tai}, {and} \bibinfo{person}{Jianmin Wang}.}
  \bibinfo{year}{2022}\natexlab{}.
\newblock \showarticletitle{Towards natural language interfaces for data
  visualization: A survey}.
\newblock \bibinfo{journal}{\emph{IEEE transactions on visualization and
  computer graphics}} (\bibinfo{year}{2022}).
\newblock


\bibitem[Siddiqui et~al\mbox{.}(2020)]%
        {siddiqui2020shapesearch}
\bibfield{author}{\bibinfo{person}{Tarique Siddiqui}, \bibinfo{person}{Paul
  Luh}, \bibinfo{person}{Zesheng Wang}, \bibinfo{person}{Karrie Karahalios},
  {and} \bibinfo{person}{Aditya Parameswaran}.}
  \bibinfo{year}{2020}\natexlab{}.
\newblock \showarticletitle{Shapesearch: A flexible and efficient system for
  shape-based exploration of trendlines}. In
  \bibinfo{booktitle}{\emph{Proceedings of the 2020 ACM SIGMOD International
  Conference on Management of Data}}. \bibinfo{pages}{51--65}.
\newblock


\bibitem[Srinivasan and Setlur(2023)]%
        {srinivasan2023bolt}
\bibfield{author}{\bibinfo{person}{Arjun Srinivasan} {and}
  \bibinfo{person}{Vidya Setlur}.} \bibinfo{year}{2023}\natexlab{}.
\newblock \showarticletitle{BOLT: A Natural Language Interface for Dashboard
  Authoring}.
\newblock  (\bibinfo{year}{2023}).
\newblock


\bibitem[Srinivasan and Stasko(2017)]%
        {srinivasan2017natural}
\bibfield{author}{\bibinfo{person}{Arjun Srinivasan} {and}
  \bibinfo{person}{John Stasko}.} \bibinfo{year}{2017}\natexlab{}.
\newblock \showarticletitle{Natural language interfaces for data analysis with
  visualization: Considering what has and could be asked}. In
  \bibinfo{booktitle}{\emph{Proceedings of the Eurographics/IEEE VGTC
  conference on visualization: Short papers}}. \bibinfo{pages}{55--59}.
\newblock


\bibitem[Vaithilingam and Guo(2019)]%
        {vaithilingam2019bespoke}
\bibfield{author}{\bibinfo{person}{Priyan Vaithilingam} {and}
  \bibinfo{person}{Philip~J Guo}.} \bibinfo{year}{2019}\natexlab{}.
\newblock \showarticletitle{Bespoke: Interactively synthesizing custom GUIs
  from command-line applications by demonstration}. In
  \bibinfo{booktitle}{\emph{Proceedings of the 32nd annual ACM symposium on
  user interface software and technology}}. \bibinfo{pages}{563--576}.
\newblock


\bibitem[Verou et~al\mbox{.}(2016)]%
        {mavo2016}
\bibfield{author}{\bibinfo{person}{Lea Verou}, \bibinfo{person}{Amy~X. Zhang},
  {and} \bibinfo{person}{David~R. Karger}.} \bibinfo{year}{2016}\natexlab{}.
\newblock \showarticletitle{Mavo: Creating Interactive Data-Driven Web
  Applications by Authoring HTML}. In \bibinfo{booktitle}{\emph{Proceedings of
  the 29th Annual Symposium on User Interface Software and Technology}} (Tokyo,
  Japan) \emph{(\bibinfo{series}{UIST '16})}. \bibinfo{publisher}{Association
  for Computing Machinery}, \bibinfo{address}{New York, NY, USA},
  \bibinfo{pages}{483–496}.
\newblock
\showISBNx{9781450341899}
\urldef\tempurl%
\url{https://doi.org/10.1145/2984511.2984551}
\showDOI{\tempurl}


\bibitem[Wang et~al\mbox{.}(2022)]%
        {wang2022towards}
\bibfield{author}{\bibinfo{person}{Yun Wang}, \bibinfo{person}{Zhitao Hou},
  \bibinfo{person}{Leixian Shen}, \bibinfo{person}{Tongshuang Wu},
  \bibinfo{person}{Jiaqi Wang}, \bibinfo{person}{He Huang},
  \bibinfo{person}{Haidong Zhang}, {and} \bibinfo{person}{Dongmei Zhang}.}
  \bibinfo{year}{2022}\natexlab{}.
\newblock \showarticletitle{Towards natural language-based visualization
  authoring}.
\newblock \bibinfo{journal}{\emph{IEEE Transactions on Visualization and
  Computer Graphics}} \bibinfo{volume}{29}, \bibinfo{number}{1}
  (\bibinfo{year}{2022}), \bibinfo{pages}{1222--1232}.
\newblock


\bibitem[Wilkinson(2005)]%
        {wilkinson2012grammar}
\bibfield{author}{\bibinfo{person}{Leland Wilkinson}.}
  \bibinfo{year}{2005}\natexlab{}.
\newblock \bibinfo{booktitle}{\emph{The Grammar of Graphics, Second Edition}}.
\newblock \bibinfo{publisher}{Springer}.
\newblock
\showISBNx{978-0-387-24544-7}


\bibitem[Wongsuphasawat et~al\mbox{.}(2017)]%
        {wongsuphasawat2017voyager}
\bibfield{author}{\bibinfo{person}{Kanit Wongsuphasawat},
  \bibinfo{person}{Zening Qu}, \bibinfo{person}{Dominik Moritz},
  \bibinfo{person}{Riley Chang}, \bibinfo{person}{Felix Ouk},
  \bibinfo{person}{Anushka Anand}, \bibinfo{person}{Jock Mackinlay},
  \bibinfo{person}{Bill Howe}, {and} \bibinfo{person}{Jeffrey Heer}.}
  \bibinfo{year}{2017}\natexlab{}.
\newblock \showarticletitle{Voyager 2: Augmenting visual analysis with partial
  view specifications}. In \bibinfo{booktitle}{\emph{Proceedings of the 2017
  chi conference on human factors in computing systems}}.
  \bibinfo{pages}{2648--2659}.
\newblock


\bibitem[Young et~al\mbox{.}(2018)]%
        {young2018recent}
\bibfield{author}{\bibinfo{person}{Tom Young}, \bibinfo{person}{Devamanyu
  Hazarika}, \bibinfo{person}{Soujanya Poria}, {and} \bibinfo{person}{Erik
  Cambria}.} \bibinfo{year}{2018}\natexlab{}.
\newblock \showarticletitle{Recent trends in deep learning based natural
  language processing}.
\newblock \bibinfo{journal}{\emph{ieee Computational intelligenCe magazine}}
  \bibinfo{volume}{13}, \bibinfo{number}{3} (\bibinfo{year}{2018}),
  \bibinfo{pages}{55--75}.
\newblock


\bibitem[Yu and Silva(2019)]%
        {yu2019flowsense}
\bibfield{author}{\bibinfo{person}{Bowen Yu} {and}
  \bibinfo{person}{Cl{\'a}udio~T Silva}.} \bibinfo{year}{2019}\natexlab{}.
\newblock \showarticletitle{FlowSense: A natural language interface for visual
  data exploration within a dataflow system}.
\newblock \bibinfo{journal}{\emph{IEEE transactions on visualization and
  computer graphics}} \bibinfo{volume}{26}, \bibinfo{number}{1}
  (\bibinfo{year}{2019}), \bibinfo{pages}{1--11}.
\newblock


\bibitem[Zamfirescu-Pereira et~al\mbox{.}(2023)]%
        {zamfirescu2023johnny}
\bibfield{author}{\bibinfo{person}{JD Zamfirescu-Pereira},
  \bibinfo{person}{Richmond~Y Wong}, \bibinfo{person}{Bjoern Hartmann}, {and}
  \bibinfo{person}{Qian Yang}.} \bibinfo{year}{2023}\natexlab{}.
\newblock \showarticletitle{Why Johnny can’t prompt: how non-AI experts try
  (and fail) to design LLM prompts}. In \bibinfo{booktitle}{\emph{Proceedings
  of the 2023 CHI Conference on Human Factors in Computing Systems}}.
  \bibinfo{pages}{1--21}.
\newblock


\bibitem[Zgraggen et~al\mbox{.}(2014)]%
        {zgraggen2014panoramicdata}
\bibfield{author}{\bibinfo{person}{Emanuel Zgraggen}, \bibinfo{person}{Robert
  Zeleznik}, {and} \bibinfo{person}{Steven~M Drucker}.}
  \bibinfo{year}{2014}\natexlab{}.
\newblock \showarticletitle{Panoramicdata: Data analysis through pen \& touch}.
\newblock \bibinfo{journal}{\emph{IEEE transactions on visualization and
  computer graphics}} \bibinfo{volume}{20}, \bibinfo{number}{12}
  (\bibinfo{year}{2014}), \bibinfo{pages}{2112--2121}.
\newblock


\bibitem[Zhang et~al\mbox{.}(2019)]%
        {zhang2019sato}
\bibfield{author}{\bibinfo{person}{Dan Zhang}, \bibinfo{person}{Yoshihiko
  Suhara}, \bibinfo{person}{Jinfeng Li}, \bibinfo{person}{Madelon Hulsebos},
  \bibinfo{person}{{\c{C}}a{\u{g}}atay Demiralp}, {and}
  \bibinfo{person}{Wang-Chiew Tan}.} \bibinfo{year}{2019}\natexlab{}.
\newblock \showarticletitle{Sato: Contextual semantic type detection in
  tables}.
\newblock \bibinfo{journal}{\emph{arXiv preprint arXiv:1911.06311}}
  (\bibinfo{year}{2019}).
\newblock


\bibitem[Zhang et~al\mbox{.}(2018)]%
        {zhang2018precision}
\bibfield{author}{\bibinfo{person}{Haoci Zhang}, \bibinfo{person}{Viraj Raj},
  \bibinfo{person}{Thibault Sellam}, {and} \bibinfo{person}{Eugene Wu}.}
  \bibinfo{year}{2018}\natexlab{}.
\newblock \showarticletitle{Precision interfaces for different modalities}. In
  \bibinfo{booktitle}{\emph{Proceedings of the 2018 International Conference on
  Management of Data}}. \bibinfo{pages}{1777--1780}.
\newblock


\end{thebibliography}

\end{document}